\documentclass[reprint,aps,superscriptaddress,physrev]{revtex4-2}
\usepackage[english]{babel}
\usepackage[table]{xcolor}
\usepackage[linktocpage=true]{hyperref}
\hypersetup{colorlinks=true,citecolor=red,urlcolor=blue}
\usepackage{amsmath}
\usepackage{amssymb}
\usepackage{braket}
\usepackage{tikz}
\usepackage{adjustbox}
\usetikzlibrary{decorations.pathmorphing}
\usetikzlibrary{arrows.meta}
\usepackage{makecell}
\usepackage{mathrsfs}  
\usepackage{graphicx}
\usepackage{colortbl}
\usepackage{dcolumn}
\usepackage[normalem]{ulem}
\usepackage{longtable}
\usepackage{longtable,booktabs}
\usepackage{multirow}
\usepackage{cancel}
\usepackage{array}  
\usepackage{bm}
\usepackage{siunitx}
\usepackage{multirow}
\usepackage{physics}
\usepackage{ltablex} 
\usepackage{ragged2e} 
\usepackage[dvipsnames]{xcolor}
\keepXColumns

\begin{document}

\preprint{}

\title{Multipole transition amplitudes and radiative decay rates in neutral cadmium}

\author{P. J. Robert}
\email{probert@unifi.it}
\affiliation{Dipartimento di Fisica e Astronomia and LENS, Università degli Studi di Firenze  \\ INFN Sezione di Firenze, Via Sansone 1, Sesto Fiorentino, Italia}
\author{S. Manzoor}
\affiliation{Dipartimento di Fisica e Astronomia and LENS, Università degli Studi di Firenze  \\ INFN Sezione di Firenze, Via Sansone 1, Sesto Fiorentino, Italia}
\author{M. Chiarotti}
\affiliation{Dipartimento di Fisica e Astronomia and LENS, Università degli Studi di Firenze  \\ INFN Sezione di Firenze, Via Sansone 1, Sesto Fiorentino, Italia}
\author{N. Poli}
\email{nicola.poli@unifi.it}
\affiliation{Dipartimento di Fisica e Astronomia and LENS, Università degli Studi di Firenze  \\ INFN Sezione di Firenze, Via Sansone 1, Sesto Fiorentino, Italia}


\begin{abstract}
We present a comprehensive study of the electronic transitions in neutral cadmium (Cd I) with a focus on forbidden transitions, motivated by recent advances in laser technology and the growing relevance of cadmium in quantum gas research, precision metrology, and atom trapping. General analytic expressions are derived for transition matrix elements of all multipolar orders, formulated to be applicable for experimental use. Using configuration interaction combined with many-body perturbation theory, we calculate not only the previously reported contributions from electric dipole (E1) transitions, but also the electric quadrupole (E2), electric octupole (E3), magnetic dipole (M1), and magnetic quadrupole transitions (M2) that have not yet been investigated for cadmium. These matrix elements are then employed to determine the lifetimes of key excited states, particularly those pertinent to laser cooling and optical frequency standards, and to evaluate the long-range dispersion coefficient $C_6$. The linewidths of the strongest transitions, along with the atomic energy levels, are compared with available experimental data to validate the accuracy of the simulations. Overall, the results are in good agreement, with the calculated energy levels exhibiting an average relative deviation of 0.3$\,\%$ from experiments. These values serve as benchmarks for both bosonic and fermionic isotopes, providing a foundation for future experimental and theoretical work in cadmium-based precision spectroscopy and cold-collision studies.

\end{abstract}

\maketitle

\section{Introduction}

The production of cold, dense, and large atomic samples is a cornerstone of modern atomic, molecular, and optical physics~\cite{Schreck2021}, underpinning a wide range of fundamental and applied experiments, including frequency metrology~\cite{Poli2013}, searches for exotic forces~\cite{Safronova2018,Sabulsky2019}, and atom interferometry~\cite{Tino2014}. State-of-the-art optical atomic clocks have reached fractional accuracies and stabilities on the order of $10^{-18}$~\cite{Brewer2019,McGrew2018,Huntemann2016,Ushijima2015,Bloom2014,Marshall2025}, motivating a potential optical redefinition of the second~\cite{LeTargat2013} and enabling applications such as chronometric leveling and searches for variations of fundamental constants~\cite{Huntemann2014,Safronova2018,Morel2020}. At this level of precision, blackbody radiation (BBR) shifts remain a significant source of systematic uncertainty~\cite{Huntemann2016,McGrew2018,Bloom2014}. While cryogenic interrogation has mitigated this effect in several systems, alternative atomic species with intrinsically low BBR sensitivity, including Hg, Mg, Tm, and Cd, offer the possibility of simpler experimental implementations with improved accuracy~\cite{McFerran2012,Yamanaka2015,Dzuba2019}.

Alkaline-earth-like atoms, such as cadmium, are particularly attractive for precision metrology due to their electronic structure, which features narrow intercombination transitions suitable for high-accuracy measurements of fundamental physics~\cite{Graham2013,Hamilton2015,Graham2016}, alongside broad, dipole-allowed transitions that enable rapid laser cooling. Advances in laser technology have produced ultra-narrow linewidth lasers~\cite{Matei2017}, making it possible to probe these forbidden transitions and to implement schemes based on magnetic-dipole couplings~\cite{Santra2005,Robert2024,He2025,Carman2025}.

Cadmium exhibits a particularly advantageous set of properties for atomic and optical applications. It possesses eight stable isotopes, most of which occur with relatively comparable natural abundances. Furthermore, the majority of its strong transitions lie in the ultraviolet domain, which makes cadmium vapor an attractive medium for ultraviolet generation via four-wave mixing~\cite{Penyazkov2025}. This feature is of particular relevance to emerging deep-UV nuclear clock schemes~\cite{Zhang2024b}.

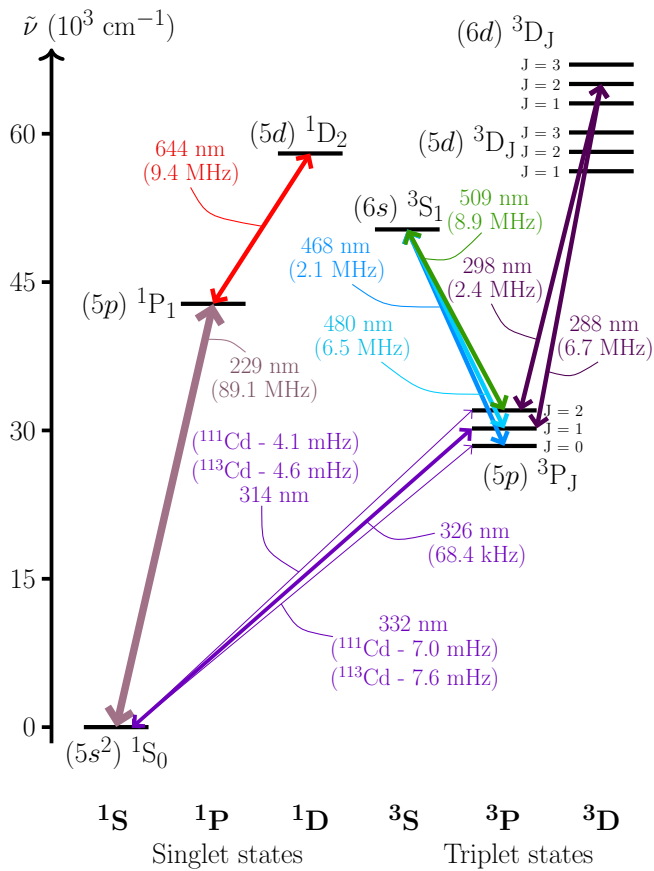
\begin{figure}[htbp] 
    \raggedright
    \begin{adjustbox}{margin*=-0.7cm 0cm 0cm 0cm}
    \newcommand{\scalelw}{0.50mm} 
    \newcommand{\firstlw}{1.00mm} 
    \newcommand{\secondlw}{0.50mm} 
    \newcommand{\repumplw}{0.60mm} 
    \newcommand{\mytextsize}{\huge} 
    \newcommand{\mybigtextsize}{\Huge}
    \newcommand{\mysmalltextsize}{\Large}
    \begin{tikzpicture}[scale=0.43, transform shape, xshift=-5cm]
  

\draw[black, ultra thick] (-1,0) -- (1,0); 
\filldraw[black] (-1.7,-0.8) circle (0pt) node[anchor=west]{\mybigtextsize $(5s^2)$~$^1$S$_0$};

\draw[black, ultra thick] (2,13.098) -- (4,13.098);
\filldraw[black] (-1.2,13.1) circle (0pt) node[anchor=west]{\mybigtextsize $(5p)$~$^1$P$_1$};

\draw[black, ultra thick] (11,8.7) -- (13,8.7);
\filldraw[black] (11.2,7.8) circle (0pt) node[anchor=west]{\mybigtextsize $(5p)$~$^3$P$_\mathrm{J}$};
\filldraw[black] (13.1,8.7) circle (0pt) node[anchor=west]{\mysmalltextsize J = 0};

\draw[black, ultra thick] (11,9.24) -- (13,9.24);
\filldraw[black] (13.1,9.25) circle (0pt) node[anchor=west]{\mysmalltextsize J = 1};

\draw[black, ultra thick] (11,9.8) -- (13,9.8);
\filldraw[black] (13.1,9.8) circle (0pt) node[anchor=west]{\mysmalltextsize J = 2};

\draw[black, ultra thick] (14,19.3) -- (16,19.3);
\filldraw[black] (10.4,21.5) circle (0pt) node[anchor=west]{\mybigtextsize $(6d)$~$^3$D$_\mathrm{J}$};
\filldraw[black] (12.4,19.3) circle (0pt) node[anchor=west]{\mysmalltextsize J = 1};

\draw[black, ultra thick] (14,19.9) -- (16,19.9);
\filldraw[black] (12.4,19.9) circle (0pt) node[anchor=west]{\mysmalltextsize J = 2};

\draw[black, ultra thick] (14,20.5) -- (16,20.5);
\filldraw[black] (12.4,20.5) circle (0pt) node[anchor=west]{\mysmalltextsize J = 3};

\draw[black, ultra thick] (14,17.2) -- (16,17.2);
\filldraw[black] (9.2,18.1) circle (0pt) node[anchor=west]{\mybigtextsize $(5d)$~$^3$D$_\mathrm{J}$};
\filldraw[black] (12.4,17.2) circle (0pt) node[anchor=west]{\mysmalltextsize J = 1};

\draw[black, ultra thick] (14,17.8) -- (16,17.8);
\filldraw[black] (12.4,17.8) circle (0pt) node[anchor=west]{\mysmalltextsize J = 2};

\draw[black, ultra thick] (14,18.4) -- (16,18.4);
\filldraw[black] (12.4,18.4) circle (0pt) node[anchor=west]{\mysmalltextsize J = 3};

\draw[black, ultra thick] (8,15.4) -- (10,15.4);
\filldraw[black] (7.2,16.1) circle (0pt) node[anchor=west]{\mybigtextsize $(6s)$~$^3$S$_1$};

\draw[black, ultra thick] (5,17.75) -- (7,17.75);
\filldraw[black] (4,18.4) circle (0pt) node[anchor=west]{\mybigtextsize $(5d)$~$^1$D$_2$};


\definecolor{myred}{RGB}{255,4,0}
\definecolor{myredtext}{RGB}{230,4,0}
\definecolor{myfirst}{RGB}{161,113,136}
\definecolor{myfirsttext}{RGB}{123,81,101} 
\definecolor{mysecond}{RGB}{109,0,188}
\definecolor{myrepumpUV}{RGB}{82,0,85}
\definecolor{my468}{RGB}{0,143,255}
\definecolor{my480}{RGB}{0,204,255}
\definecolor{my480text}{RGB}{0,184,230}
\definecolor{my509}{RGB}{51,153,0}

\draw[myfirst, line width=\firstlw, {Straight Barb[length=1.7mm]}-{Straight Barb[length=1.7mm]}] (0,0) -- (3,13.098);
\draw[myfirst] (2.7,12) .. controls (3.5,12) .. (4,11.6);
\filldraw[myfirsttext] (3.4,11.2) circle (0pt) node[anchor=west]{\mytextsize 229 nm};
\filldraw[myfirsttext] (2.8,10.4) circle (0pt) node[anchor=west]{\mytextsize (89.1 MHz)};

\draw[mysecond, line width=\secondlw, {Straight Barb[length=1mm]}-{Straight Barb[length=1mm]}] (0.5,0) -- (11,9.24);
\draw[mysecond] (7.7,6.4) .. controls (8.5,5.7) .. (9.85,6);
\filldraw[mysecond] (9.9,6.1) circle (0pt) node[anchor=west]{\mytextsize 326 nm};
\filldraw[mysecond] (9.5,5.3) circle (0pt) node[anchor=west]{\mytextsize (68.4 kHz)};

\draw[mysecond, line width=0.00mm, {Straight Barb[length=0.8mm]}-{Straight Barb[length=0.8mm]}] (0.5,0) -- (11,8.7);
\draw[mysecond] (5.1,3.82) .. controls (5.6,3) .. (6.7,2.8);
\filldraw[mysecond] (8,3.2) circle (0pt) node[anchor=west]{\mytextsize 332 nm};
\filldraw[mysecond] (6.5,2.4) circle (0pt) node[anchor=west]{\mytextsize ($^{111}\mathrm{Cd}$  -  7.0 mHz)};
\filldraw[mysecond] (6.5,1.5) circle (0pt) node[anchor=west]{\mytextsize ($^{113}\mathrm{Cd}$ - 7.6 mHz)};

\draw[mysecond, line width=0.00mm, {Straight Barb[length=0.8mm]}-{Straight Barb[length=0.8mm]}] (0.5,0) -- (11,9.8);
\draw[mysecond] (5.65,4.8) .. controls (4.5,5.5) .. (4.5,6.85);
\filldraw[mysecond] (3.7,7.2) circle (0pt) node[anchor=west]{\mytextsize 314 nm};
\filldraw[mysecond] (2.2,8.8) circle (0pt) node[anchor=west]{\mytextsize ($^{111}\mathrm{Cd}$  - 4.1 mHz)};
\filldraw[mysecond] (2.2,7.9) circle (0pt) node[anchor=west]{\mytextsize ($^{113}\mathrm{Cd}$  - 4.6 mHz)};

\draw[myred, line width=\repumplw, {Straight Barb[length=1.2mm]}-{Straight Barb[length=1.2mm]}] (3,13.098) -- (6,17.75);
\draw[myred] (4.5,15.5) .. controls (2.8,15.8) .. (2.3,16.7);
\filldraw[myredtext] (1.1,17.9) circle (0pt) node[anchor=west]{\mytextsize 644 nm};
\filldraw[myredtext] (0.7,17.1) circle (0pt) node[anchor=west]{\mytextsize (9.4 MHz)};

\draw[my468, line width=\repumplw, {Straight Barb[length=1.2mm]}-{Straight Barb[length=1.2mm]}] (12,8.7) -- (9,15.4);
\draw[my468] (10,13) .. controls (8.3,13) .. (7.3,13.7);
\filldraw[my468] (5.6,14.9) circle (0pt) node[anchor=west]{\mytextsize 468 nm};
\filldraw[my468] (5.2,14.1) circle (0pt) node[anchor=west]{\mytextsize (2.1 MHz)};

\draw[my480, line width=\repumplw, {Straight Barb[length=1.2mm]}-{Straight Barb[length=1.2mm]}] (12,9.24) -- (9,15.4);
\draw[my480] (11.5,10.3) .. controls (9.5,10.3) .. (8.2,11.3);
\filldraw[my480text] (6.3,12.5) circle (0pt) node[anchor=west]{\mytextsize 480 nm};
\filldraw[my480text] (6,11.7) circle (0pt) node[anchor=west]{\mytextsize (6.5 MHz)};

\draw[my509, line width=\repumplw, {Straight Barb[length=1.2mm]}-{Straight Barb[length=1.2mm]}] (12,9.8) -- (9,15.4);
\draw[my509] (9.5,14.5) .. controls (10.4,14.4) .. (11.3,15.3);
\filldraw[my509] (10.5,16.4) circle (0pt) node[anchor=west]{\mytextsize 509 nm};
\filldraw[my509] (10.1,15.6) circle (0pt) node[anchor=west]{\mytextsize (8.9 MHz)};


\draw[myrepumpUV, line width=\repumplw, {Straight Barb[length=1.2mm]}-{Straight Barb[length=1.2mm]}] (12.5,9.8) -- (15,19.9);
\draw[myrepumpUV] (13,12) .. controls (12,12.3) .. (11.6,13.1);
\filldraw[myrepumpUV] (10.6,14.3) circle (0pt) node[anchor=west]{\mytextsize 298 nm};
\filldraw[myrepumpUV] (10.1,13.5) circle (0pt) node[anchor=west]{\mytextsize (2.4 MHz)};

\draw[myrepumpUV, line width=\repumplw, {Straight Barb[length=1.2mm]}-{Straight Barb[length=1.2mm]}] (13,9.24) -- (15,19.9);
\draw[myrepumpUV] (13.35,11) .. controls (14.1,10.8) .. (14.9,11.3);
\filldraw[myrepumpUV] (13.9,12.5) circle (0pt) node[anchor=west]{\mytextsize 288 nm};
\filldraw[myrepumpUV] (13.5,11.7) circle (0pt) node[anchor=west]{\mytextsize (6.7 MHz)};



\filldraw[black] (1,-4) circle (0pt) node[anchor=west]{\mybigtextsize Singlet states};
\filldraw[black] (-0.7,-2.8) circle (0pt) node[anchor=west]{\mybigtextsize $\mathbf{^1S}$};
\filldraw[black] (2.3,-2.8) circle (0pt) node[anchor=west]{\mybigtextsize $\mathbf{^1P}$};
\filldraw[black] (5.3,-2.8) circle (0pt) node[anchor=west]{\mybigtextsize $\mathbf{^1D}$};

\filldraw[black] (10,-4) circle (0pt) node[anchor=west]{\mybigtextsize Triplet states};
\filldraw[black] (8.3,-2.8) circle (0pt) node[anchor=west]{\mybigtextsize $\mathbf{^3S}$};
\filldraw[black] (11.3,-2.8) circle (0pt) node[anchor=west]{\mybigtextsize $\mathbf{^3P}$};
\filldraw[black] (14.3,-2.8) circle (0pt) node[anchor=west]{\mybigtextsize $\mathbf{^3D}$};

\draw[line width=\scalelw, ->] (-2,-0.5) -- (-2,21);
\filldraw[black] (-3,21.7) circle (0pt) node[anchor=west]{\mybigtextsize $\tilde{\nu}$ (10$^3$ cm$^{-1}$)};

\draw[line width=\scalelw] (-1.97,0) -- (-2.3,0);
\filldraw[black] (-3,0) circle (0pt) node[anchor=west]{\mybigtextsize 0};

\draw[line width=\scalelw] (-1.97,4.59) -- (-2.3,4.59);
\filldraw[black] (-3.4,4.59) circle (0pt) node[anchor=west]{\mybigtextsize 15};

\draw[line width=\scalelw] (-1.97,9.18) -- (-2.3,9.18);
\filldraw[black] (-3.4,9.18) circle (0pt) node[anchor=west]{\mybigtextsize 30};

\draw[line width=\scalelw] (-1.97,13.77) -- (-2.3,13.77);
\filldraw[black] (-3.4,13.77) circle (0pt) node[anchor=west]{\mybigtextsize 45};

\draw[line width=\scalelw] (-1.97,18.36) -- (-2.3,18.36);
\filldraw[black] (-3.4,18.36) circle (0pt) node[anchor=west]{\mybigtextsize 60};

\end{tikzpicture}
\end{adjustbox}
\caption{\justifying Simplified energy-level structure of neutral cadmium (Cd I) and its principal optical transitions relevant to laser-cooling and spectroscopic applications. Wavelenghs and natural linewidths $\Gamma/2\pi$ of the transitions are also reported. The different $^3$D$_\mathrm{J}$ and $^3$P$_\mathrm{J}$ levels are not drawn to scale; they have been separated for readability.}
\label{fig:levelStructure}
\end{figure}


These characteristics make it an excellent candidate for probing physics beyond the Standard Model through the measurement of isotope shifts using techniques such as King’s plots~\cite{Hur2022, Ohayon2022}. The bosonic isotopes allow access to ultranarrow clock transitions, while the fermionic isotopes possess nuclear spin, providing hyperfine-induced clock transitions with natural linewidths of $\Gamma/2\pi = 7.0$ and $7.6$\,mHz~\cite{Garstang1962}. Figure \ref{fig:levelStructure} presents the most common transitions of neutral cadmium employed in cold atom physics.

The spin-forbidden $(5s^2)$~$^1$S$_0$ $\rightarrow$ $(5s5p)$~$^3$P$_1$ transition at 326\,nm enables Doppler cooling to the $\mu$K level~\cite{Brickman2007,Bandarupally2023,PadillaCastilloarxiv}. Its short wavelength also reduces radiation trapping and allows the preparation of dense, cold ensembles, which may enable rapid or even continuous production of quantum-degenerate gases~\cite{Walker1990,Duarte2011,Stellmer2013,Chen2022}.

Despite these attractive features, experimental realizations of cold Cd remain limited. Early demonstrations include magneto-optical traps (MOTs) on the broad {$(5s^2)$~$^1$S$_0$ $\rightarrow (5s5p)$~$^1$P$_1$} transition at 229\,nm~\cite{Brickman2007} and, more recently, on the narrow 326\,nm intercombination line~\cite{Kaneda2016,Yamaguchi2019}. The development of robust, high-flux sources is hindered by challenges associated with ultraviolet light generation, vacuum compatibility, and photoionization~\cite{Tinsley2022,PadillaCastilloarxiv}.

Most studies on cadmium have focused on a limited number of transitions, primarily restricted to low-lying states~\cite{Moszynski2003,Dzuba2019,Yamaguchi2019,Porsev2020}. While more recent works employing Configuration Interaction combined with Many-Body Perturbation Theory (CI+MBPT) have extended the scope to larger sets of transitions, including higher-lying states~\cite{Zhang2024, Penyazkov2025}, these investigations have remained limited to the so-called ‘allowed’ electric-dipole transitions. Complementary studies based on Multiconfiguration Dirac-Hartree-Fock combined with Configuration Interaction have investigated cadmium isotope shifts and clock-transition frequencies, including detailed King-plot analyses for the extraction of nuclear parameters~\cite{Schelfhout2022}. Nevertheless, systematic calculations of higher-order forbidden multipole transitions remain lacking.

In this work, we compute, using CI+MBPT, the previously unreported higher-order multipole matrix elements E2, E3, M1, and M2 of neutral cadmium. These matrix elements are subsequently employed to analyze the lifetimes of key transitions with particular emphasis on the previously unexplored bosonic clock transition and to determine the long-range van der Waals $C_6$ coefficient. Unless otherwise specified, all formulas and results are expressed in atomic units.

\section{CI+MBPT framework}

\subsection{Theory}
CI+MBPT is an ab initio approach that has shown high precision in calculating atomic properties~\cite{Berengut2016,Torretti2017,Berengut2006}. This approach combines Configuration Interaction (CI) to account for valence–valence electron correlations, with Many-Body Perturbation Theory (MBPT) to treat core–valence interactions~\cite{Dzuba1996}.

The procedure begins by solving the Dirac-Fock (DF) equations, the relativistic generalization of the Hartree-Fock equations, for both core and valence electrons. This is done within one of the standard approximations of the potential: $V^N$, $V^{N-1}$, or $V^{N-M}$, where $N$ is the total number of electrons and $M$ is the number of valence electrons. The DF Hamiltonian takes the form~\cite{johnson2007atomic}
\begin{equation}
\label{eq1}
h_{\mathrm{DF}} = c\,\boldsymbol{\alpha} \cdot \mathbf{p} + (\beta - 1)\,c^{2} - \frac{Z}{r} + V^{N_{\mathrm{DF}}}
\end{equation}
where $\boldsymbol{\alpha}$ and $\beta$ are the Dirac matrices and $V^{N_{\mathrm{DF}}}$ is the self-consistent potential in the chosen approximation.

The one-electron solutions of the Dirac equation can be expressed as
\begin{equation}
\label{eq2}
\psi (r) = \frac{1}{r}
\begin{pmatrix}
iP_{n\kappa}(r)\,\Omega_{\kappa m}(\hat{r}) \\
Q_{n\kappa}(r)\,\Omega_{-\kappa m}(\hat{r})
\end{pmatrix}
\end{equation}
where $\kappa = (-1)^{j+l+\frac{1}{2}}(j+\frac{1}{2})$, $P_{n\kappa}$ and $Q_{n\kappa}$ typically refer to the large and small radial components respectively and $\Omega_{\kappa m}$ are the spherical spinors.

The DF operator can be extended to include relativistic corrections such as the Breit interaction and QED effects~\cite{johnson2007atomic}, with the Breit operator given by
\begin{equation}
\label{eq3}
B_{ij} = -\frac{1}{2\,r_{ij}}
\left( \boldsymbol{\alpha}_i\cdot\boldsymbol{\alpha}_j+ \frac{(\boldsymbol{\alpha}_i\cdot\boldsymbol{r}_{ij})\,(\boldsymbol{\alpha}_j\cdot\boldsymbol{r}_{ij})}{r^2_{ij}} \right)
\end{equation}

The remaining valence orbitals and pseudostates are constructed as linear combinations of B-spline basis functions~\cite{Kahl2019}. The resulting one-particle basis functions are then used to build anti-symmetrized many-electron basis states $\ket{\text{proj}_n}$, expressed as superpositions of Slater determinants~\cite{johnson2007atomic}.

Configuration state functions (CSFs) $\ket{I}$ are defined as eigenfunctions of $\hat{J}^2$ and $\hat{J}z$, obtained as linear combinations of $\ket{\text{proj}_n}$ :

\begin{equation}
\label{eq4}
\ket{I} = \sum_{n} c_n \ket{\text{proj}_n}
\end{equation}

The atomic wavefunction is expressed as a CI expansion over the CSFs~\cite{Kahl2019}:

\begin{equation}
\label{eq4}
\Psi = \sum_{I\in P} C_I \ket{I}
\end{equation}

where $P$ denotes the model space included in the CI calculation.

The dimension of the CI matrix grows rapidly with the number of orbitals included, making it computationally prohibitive to account for all core–valence correlations or core excitations directly within CI. These effects are instead incorporated perturbatively: MBPT is applied to second order to modify the CI matrix elements ~\cite{Dzuba1996,Berengut2006}. The resulting CI+MBPT eigenvalue problem is
\begin{equation}
\label{eq5}
\sum_{J \in P} \left( H_{IJ}+\sum_{M \in Q} \frac{\bra{I} \hat{H} \ket{M} \bra{M} \hat{H} \ket{J}}{E - E_M} \right) C_J = E\,C_I
\end{equation}
where $Q$ is the complementary space containing states excluded from the CI model space $P$.

\subsection{AMBiT}
AMBiT is a software for fully relativistic atomic structure calculations, integrating CI+MBPT to enable determinations of energy levels, transition matrix elements, g-factors, and isotope shifts in complex systems.
In \textsc{AMBiT}, the CI space is assembled from CSFs generated by permitting electron and/or hole excitations from a designated set~\cite{Kahl2019}. The allowed excitations are restricted by upper bounds on the principal quantum number $n$ and orbital angular momentum $l$. 
To further decrease the computational burden, \textsc{AMBiT} employs the \emph{emu CI} strategy~\cite{Geddes2018}. This approach partitions the CSFs into two groups: the \emph{large-side} set, constructed with more generous limits on $n$ and $l$, and the \emph{small-side} set, produced with tighter restrictions.
Frequently, the small-side space is supplemented by configurations obtained through single excitations to higher-$n$ and higher-$l$ orbitals than those defining its primary limits. Core–valence correlations are subsequently incorporated in a perturbative fashion via the MBPT operator, contributing second-order energy shifts.

The choice of parameters for the cadmium simulations with AMBiT is guided by the work of Penyazkov \textit{et al.}~\cite{Penyazkov2025}, who employed AMBiT for the calculation of E1 matrix elements. However, whereas most studies using CI+MBPT have focused on specific states for targeted applications, such as four-wave mixing~\cite{Penyazkov2025} or magic trapping conditions~\cite{Zhang2024}, our objective is to minimize the overall relative deviation between calculated and experimental energy levels~\cite{Burns1956,Brown1975,VidolovaAngelova1996,Moore1971,NIST_ASD} rather than optimize for a restricted set of states. Consequently, while our approach is inspired by these earlier works, some differences in computed values arise. The parameters were iteratively optimized until the mean relative deviation between the computed energy levels and those reported in the NIST database was reduced to below 0.5\,$\%$, with a maximum relative deviation of 5\,$\%$.

The final computational setup employed a $V^{N-2}$ potential to solve the Dirac–Hartree–Fock (DHF) equations. The valence orbital basis is specified as $12spdf$, including all $s$-, $p$-, $d$-, and $f$-orbitals with $n<12$. The $5s^2$ and $5s,5p$ configurations serve as the reference set. Large-side CSFs are obtained by allowing all single and double excitations from these references into orbitals up to $12spdf$. The small-side CSFs are generated from the same references, with single and double excitations restricted to $6sp5d4\!f$, in addition to single excitations extending up to $20spdf$. In every calculation, single and double excitations from the $4d$ core are also included. MBPT corrections are computed by accounting for all one-, two-, and three-body diagrams within a $30spdf\!g$ basis. The present calculation does not include ionization effects.

\section{Reduced matrix elements}
\subsection{Theory and selection rules}
This section is largely based on the lecture notes of W. R. Johnson~\cite{johnson2007atomic}. For more detailed discussions, particularly regarding selection rules, the reader can refer to Refs.~\cite{Sobelman1979,harris1978symmetry,foot2005atomic}.

Let $\psi(\boldsymbol{r})$ denote the Dirac wavefunction, $\hat{\epsilon}$ the polarization vector of the electromagnetic field, $\boldsymbol{k}$ its wave vector, and $\ket{a}$ and $\ket{b}$ the initial and final states, respectively. The transition amplitude for a one-electron transition can be expressed as
\begin{equation}
\label{eq6}
    T_{ba} = \int \psi_b^\dagger(r)\,\boldsymbol{\alpha}\cdot \hat \epsilon \,e^{i\boldsymbol{k}\cdot\boldsymbol{r}}\psi_a(r)\,d^3r
\end{equation}

The exponential factor $e^{i\boldsymbol{k}\cdot\boldsymbol{r}}$ is of central interest, as it generates the multipole operators. For neutral atoms or ions with small nuclear charge $Z$, the condition $|\boldsymbol{k}\cdot\boldsymbol{r}| \ll 1$ holds, allowing the exponential to be expanded in a Taylor series.

To zeroth order, the exponential term reduces to unity, and within the Pauli approximation, the nonrelativistic length-form transition amplitude is obtained
\begin{equation}
\label{eq7}
    T_{ba}^{(0)} = i\,\frac{\omega_{ba}}{c}\bra{b}\boldsymbol{r}\ket{a}\cdot \hat \epsilon
\end{equation}
where $\hbar\,\omega_{ba}$ is the energy difference between $\ket{a}$ and $\ket{b}$.

Since this amplitude is proportional to the electric-dipole operator $\boldsymbol{d} = e\, \boldsymbol{r}$, it is referred to as the electric-dipole (E1) amplitude. In the spherical basis, the electric-dipole operator is an odd-parity irreducible tensor of rank one. Consequently, $\ket{a}$ and $\ket{b}$ must have opposite parity for the transition matrix element to be nonzero, representing a fundamental selection rule for E1 transitions.

The next-order term in the nonrelativistic expansion, $i\, \boldsymbol{k}\cdot\boldsymbol{r}$, gives rise to the transition amplitude

\begin{equation}
\label{eq8}
    T_{ba}^{(1)} = i\,\frac{k}{c}\bra{b}\boldsymbol{M}\ket{a}\cdot\left[\hat k \times \hat \epsilon\right] - \frac{k\,\omega_{ba}}{6\,c}\sum_{i,j}\bra{b}Q_{ij}\ket{a}\hat k_i\,\hat \epsilon_j
\end{equation}
where $\boldsymbol{M} = \frac{1}{2 m} (\boldsymbol{L} + 2 \boldsymbol{S})$ is the magnetic-dipole operator, and $Q{ij} = 3 x_i x_j - r^2 \delta{ij}$ is the electric quadrupole operator. The first term corresponds to the magnetic-dipole (M1) amplitude, while the second term represents the electric quadrupole (E2) amplitude.

In general, under the long-wavelength approximation and the Pauli approximation, the one-electron electric multipole reduced matrix elements of order $\eta$ in the \textit{length gauge} can be written as

\begin{equation}
\label{eq9}
\begin{split}
\bra{\kappa_a}\boldsymbol{O}^{(E\eta)}\ket{\kappa_b} ={} & 
\bra{\kappa_a}C_{\eta}\ket{\kappa_b} \\
& \times \int_0^{\infty} r^{\eta} \left[ P_a(r)\,P_b(r) + Q_a(r)\,Q_b(r) \right]\,dr
\end{split}
\end{equation}

while the magnetic multipole reduced matrix elements of order $\eta$ in the \textit{transverse gauge} are given by
\begin{equation}
\label{eq10}
\begin{split}
    \bra{\kappa_a}\boldsymbol{O}^{(M\eta)}\ket{\kappa_b} = {} &
    \bra{-\kappa_a}C_{\eta}\ket{\kappa_b}
    \frac{\kappa_a+\kappa_b} {\eta+1}\\ & \times\int_0^{\infty}r^{\eta}\left[P_a(r)\,Q_b(r)+Q_a(r)\,P_b(r)\right]\,dr
    \end{split}
\end{equation}

where $C_{\lambda}$ denotes the normalized spherical harmonic tensor, and the approximation $j_n(z) \sim z^n / (2n + 1)!!$ has been used for the spherical Bessel functions of order $n$~\cite{NIST_DLMF,abramowitz_stegun_1964}. While Eqs.~\ref{eq7}–\ref{eq10} are written in the nonrelativistic form, AMBiT instead performs its simulations using the full Dirac equation and therefore evaluates relativistic matrix elements. The general matrix element is then obtained via the Wigner-Eckart Theorem.

Focusing on matrix elements up to E3 and M2, the corresponding selection rules are summarized in Table \ref{tab:selection_rules}.

\begin{table}[hbtp]
\caption{\justifying Selection rules for electric (E1–E3) and magnetic (M1–M2) multipole transitions, showing allowed changes in total angular momentum $J$, parity, and the corresponding restrictions in $LS$ coupling.}
\label{tab:selection_rules}
\begin{tabular}{c|c|c|ccc} 
\hline\hline
Type & $\Delta J$ & \makecell{Parity \\ Change} & \multicolumn{2}{c}{In $LS$ coupling} \\ \hline
\makecell{E1} & \makecell{$0, \pm 1$\\(0 $\not\to$ 0)} & \makecell{Yes} & \makecell{$\Delta S = 0$} & \makecell{$\Delta L = 0,\pm 1$\\(0$\not\to$0)} \\ \hline 
\makecell{E2} & \makecell{$0, \pm 1, \pm 2$\\(0 $\not\to$ 0,1)\\($\frac{1}{2} \not\to \frac{1}{2}$)} & \makecell{No} & \makecell{$\Delta S = 0$} & \makecell{$\Delta L=0,\pm 1,\pm 2$ \\($0\not\to0,1$)} \\ \hline
\makecell{E3} & \makecell{$0, \pm 1, \pm 2, \pm 3$\\(0 $\not\to$ 0,1,2)\\($\frac{1}{2} \not\to \frac{1}{2},\frac{3}{2}$)\\(1$\not\to$1)} & \makecell{Yes} & \makecell{$\Delta S = 0$} &\makecell{$\Delta L=0, \pm 1,\pm 2, \pm 3$\\(0$\not\to$0,1,2)\\(1$\not\to$1)} \\ \hline
\makecell{M1} & \makecell{$0, \pm 1$\\(0 $\not\to$ 0)} & \makecell{No} & \makecell{$\Delta S = 0$} & \makecell{$\Delta L=0$} \\ \hline
\makecell{M2} & \makecell{$0, \pm 1, \pm 2$\\(0 $\not\to$ 0,1)\\($\frac{1}{2} \not\to \frac{1}{2}$)} & \makecell{Yes} & \makecell{$\Delta S = 0$} & \makecell{$\Delta L=0,\pm 1,\pm 2$ \\($0\not\to0,1$)} \\
\hline\hline
\end{tabular}
\end{table}

\subsection{Level Structure}

The calculated energy levels for states up to the ground state of the 9$^{\mathrm{th}}$ shell are presented in Table \ref{tab:level_structure}. Each level was cross-checked accounting for the relativistic {$g\text{-}$factor}, and the corresponding magnetic moment amplitude, $|\mu| =|g\,\mu_B\,\textbf{J}/\hbar\,|= g\,\mu_B \sqrt{J\,(J+1)}$, is reported in units of the Bohr magneton with $\textbf{J}$ the total angular momentum operator and $J$ its associated quantum number~\cite{AtkinsFriedman2010}.

The AMBiT parameters were optimized to reproduce the experimental energy levels reported in the NIST database~\cite{Burns1956,Brown1975,VidolovaAngelova1996,Moore1971,NIST_ASD}, with an overall relative deviation below 0.5\,$\%$ and a maximum relative deviation below 5\,$\%$. In the final configuration (Table~\ref{tab:level_structure}), the obtained average relative deviation is 0.3\,$\%$, with the largest relative deviation of 1.4\,$\%$ occurring for the highest calculated state, $(5s9s)$~$^1$S$_0$. The computation time was on the order of 10 hours using a standard workstation with 48\,GB of RAM. This comparison with experimental data serves to ensure the highest accuracy of the subsequent reduced matrix elements.

\begin{table}[htbp]
\caption{\justifying Energy levels (in cm$^{-1}$), $g$-factors, and magnetic moment amplitudes (in $\mu_B$) 
for neutral cadmium states from ($5s^2$)~$^1$S$_0$ to ($5s9s$)~$^1$S$_0$. 
CI+MBPT results are compared with the NIST database. The atomic states are represented by symbols of the form $^{2\text{S}+1}$L$^{\text{parity}}_\text{J}$.}
\label{tab:level_structure}
\rowcolors{2}{gray!10}{white}
\begin{ruledtabular}
\begin{tabular}{ccccl}
State           & $g$-factor & $|\mu|$   & CI+MBPT     & \multicolumn{1}{c}{NIST}        \\
                &            & ($\mu_B$) & (cm$^{-1}$) & \multicolumn{1}{c}{(cm$^{-1}$)} \\ \hline \addlinespace
$(5s^2)$~$^1$S$_0$     & /       & 0      &  0.000 &  \multicolumn{1}{>{\color{black}}r}{0.000 (0.007)} \\ 
$(5s5p)$~$^3$P$_0^o$  & /       & 0      & 30 440 & 30 113.990 (0.002) \\ 
$(5s5p)$~$^3$P$^o_1$  & 1.4987  & 2.1195 & 30 998 & 30 656.087 (0.002) \\ 
$(5s5p)$~$^3$P$_2^o$  & 1.5000  & 3.6742 & 32 200 & 31 826.952 (0.002) \\ 
$(5s5p)$~$^1$P$^o_1$  & 1.0012  & 1.4159 & 44 076 & 43 692.384 (0.002) \\ 
$(5s6s)$~$^3$S$_1$    & 2.0000  & 2.8284 & 51 477 & 51 483.980 (0.002) \\ 
$(5s6s)$~$^1$S$_0$    & /       & 0      & 53 461 & 53 310.101 (0.010) \\ 
$(5s6p)$~$^3$P$^o_0$  & /       & 0      & 58 300 & 58 390.9 \ \ \ (2.5) \\ 
$(5s6p)$~$^3$P$^o_1$  & 1.4974  & 2.1176 & 58 373 & 58 461.6 \ \ \ (2.5) \\ 
$(5s6p)$~$^3$P$^o_2$  & 1.5000  & 3.6742 & 58 548 & 58 635.7 \ \ \ (2.5) \\ 
$(5s5d)$~$^1$D$_2$    & 1.0001  & 2.4500 & 59 205 & 59 219.734 (0.002) \\ 
$(5s5d)$~$^3$D$_1$    & 0.5000  & 0.7071 & 59 396 & 59 485.768 (0.002) \\ 
$(5s5d)$~$^3$D$_2$    & 1.1666  & 2.8576 & 59 409 & 59 497.868 (0.002) \\ 
$(5s5d)$~$^3$D$_3$    & 1.3333  & 4.6187 & 59 430 & 59 515.990 (0.020) \\ 
$(5s6p)$~$^1$P$^o_1$  & 1.0026  & 1.4179 & 60 099 & 59 907.28 \, (0.18) \\ 
$(5s7s)$~$^3$S$_1$    & 2.0000  & 2.8284 & 62 415 & 62 563.435   (0.002) \\ 
$(5s7s)$~$^1$S$_0$    & /       & 0      & 62 984 & 63 086.896   (0.002) \\ 
$(5s7p)$~$^3$P$^o_0$  & /       & 0      & 64 870 & 64 995.9 \ \ \ (2.5) \\ 
$(5s7p)$~$^3$P$^o_1$  & 1.4957  & 2.1152 & 64 902 & 65 025.5 \ \ \ (0.3) \\ 
$(5s7p)$~$^3$P$^o_2$  & 1.5000  & 3.6742 & 64 971 & 65 093.702 (0.016) \\ 
$(5s6d)$~$^1$D$_2$    & 1.0001  & 2.4497 & 65 068 & 65 134.783 (0.002) \\ 
$(5s6d)$~$^3$D$_1$    & 0.5000  & 0.7071 & 65 234 & 65 353.372 (0.002) \\ 
$(5s6d)$~$^3$D$_2$    & 1.1667  & 2.8578 & 65 240 & 65 358.881 (0.002) \\ 
$(5s6d)$~$^3$D$_3$    & 1.3333  & 4.6187 & 65 250 & 65 367.227 (0.020) \\ 
$(5s7p)$~$^1$P$^o_1$  & 1.0043  & 1.4202 & 65 447 & 65 501.412 (0.016) \\ 
$(5s4f)$~$^3$F$^o_2$  & 0.6667  & 1.6330 & 65 448 & 65 586.0 \ \ \ (0.3) \\ 
$(5s4f)$~$^3$F$^o_3$  & 1.0668  & 3.6955 & 65 448 & 65 586.0 \ \ \ (0.3) \\ 
$(5s4f)$~$^3$F$^o_4$  & 1.2500  & 5.5902 & 65 450 & 65 586.0 \ \ \ (0.3) \\
$(5s8s)$~$^3$S$_1$    & 2.0000  & 2.8284 & 66 576 & 66 682.029 (0.002) \\ 
$(5s8s)$~$^1$S$_0$    & /       & 0      & 66 849 & 66 905.641 (0.010) \\ 
$(5s8p)$~$^3$P$^o_0$  & /       & 0      & 67 940 & 67 829.656 (0.020) \\ 
$(5s8p)$~$^3$P$^o_1$  & 1.4949  & 2.1141 & 67 960 & 67 842.06 \, (0.05) \\ 
$(5s8p)$~$^3$P$^o_2$  & 1.5000  & 3.6742 & 68 008 & 67 875.191 (0.016) \\ 
$(5s7d)$~$^1$D$_2$    & 1.0001  & 2.4497 & 67 910 & 67 838.401 (0.002) \\ 
$(5s7d)$~$^3$D$_1$    & 0.5000  & 0.7071 & 68 073 & 67 989.814 (0.002)  \\ 
$(5s7d)$~$^3$D$_2$    & 1.1667  & 2.8578 & 68 077 & 67 992.708 (0.002)  \\ 
$(5s7d)$~$^3$D$_3$    & 1.3333  & 4.6186 & 68 080 & 67 997.101 (0.020) \\ 
$(5s8p)$~$^1$P$^o_1$  & 1.0051  & 1.4214 & 68 323 & 68 059.393 (0.016) \\ 
$(5s5f)$~$^3$F$^o_2$  & 0.6667  & 1.6330 & 68 088 & 68 093.7 \ \ \ (2.5) \\ 
$(5s5f)$~$^3$F$^o_3$  & 1.0441  & 3.6169 & 68 088 & 68 093.7 \ \ \ (2.5) \\ 
$(5s5f)$~$^3$F$^o_4$  & 1.2500  & 5.5902 & 68 090 & 68 093.7 \ \ \ (2.5) \\ 
$(5s9s)$~$^3$S$_1$    & 2.0000  & 2.8284 & 69 488 & 68 682.325 (0.002) \\ 
$(5s9s)$~$^1$S$_0$    & /       & 0      & 69 788 & 68 798.760 (0.010) \\ 
\end{tabular}
\end{ruledtabular}
\end{table}

\begin{table*}[hbtp!]
\centering
\rowcolors{2}{gray!10}{white}

\begin{ruledtabular}
\caption{\justifying Matrix elements in atomic units for electric dipole, electric quadrupole, electric octupole, magnetic dipole, and magnetic quadrupole transitions. The corresponding wavelengths in ångström are taken from the NIST database.}

\label{tab:mat_elem}
\begin{tabular}{ccccccc} 
States & $\lambda$ ($\SI{}{\angstrom}$) & E1        & E2        & E3        & M1        & M2 \\ \hline \addlinespace
$(5s^2)$~$^1$S$_0$ $\leftrightarrow$ $(5s5p)$~$^3$P$^o_1$ &3262& 1.784$\times$10$^{-1}$ & & & & \\ 
 $(5s^2)$~$^1$S$_0$ $\leftrightarrow$ $(5s5p)$~$^3$P$^o_2$ &3142& & & & & 1.497$\times$10$^{1}$ \\ 
$(5s^2)$~$^1$S$_0$ $\leftrightarrow$ $(5s5p)$~$^1$P$^o_1$ &2289& 3.788 & & & & \\ 
$(5s^2)$~$^1$S$_0$ $\leftrightarrow$ $(5s5d)$~$^1$D$_2$ &1689& & 1.201$\times$10$^{1}$ & & & \\ 
$(5s^2)$~$^1$S$_0$ $\leftrightarrow$ $(5s5d)$~$^3$D$_2$ &1681& & 2.601$\times$10$^{-1}$ & & & \\ 
$(5s^2)$~$^1$S$_0$ $\leftrightarrow$ $(5s6p)$~$^1$P$^o_1$ &1669& 9.096$\times$10$^{-1}$ & & & & \\ 
$(5s^2)$~$^1$S$_0$ $\leftrightarrow$ $(5s6d)$~$^1$D$_2$ &1535& & 6.203 & & & \\ 
$(5s^2)$~$^1$S$_0$ $\leftrightarrow$ $(5s6d)$~$^3$D$_2$ &1530& & 5.204$\times$10$^{-2}$ & & & \\ 
$(5s^2)$~$^1$S$_0$ $\leftrightarrow$ $(5s7p)$~$^1$P$^o_1$ &1527& 4.643$\times$10$^{-1}$ & & & & \\ 
$(5s^2)$~$^1$S$_0$ $\leftrightarrow$ $(5s8p)$~$^1$P$^o_1$ &1469& 3.563$\times$10$^{-1}$ & & & & \\ 
$(5s5p)$~$^3$P$^o_0$ $\leftrightarrow$ $(5s5p)$~$^3$P$^o_1$ &1.845$\times$10$^{5}$& & & & 1.412 & \\ 
$(5s5p)$~$^3$P$^o_0$ $\leftrightarrow$ $(5s5p)$~$^3$P$^o_2$ &5.838$\times$10$^{4}$& & 1.205$\times$10$^{1}$ & & & \\ 
$(5s5p)$~$^3$P$^o_0$ $\leftrightarrow$ $(5s5p)$~$^1$P$^o_1$ &7365& & & & 6.155$\times$10$^{-2}$ & \\ 
$(5s5p)$~$^3$P$^o_0$ $\leftrightarrow$ $(5s6s)$~$^3$S$_1$ &4679& 1.479 & & & & \\ 
$(5s5p)$~$^3$P$^o_0$ $\leftrightarrow$ $(5s5d)$~$^1$D$_2$ &3436& & & & & 7.581 \\ 
$(5s5p)$~$^3$P$^o_0$ $\leftrightarrow$ $(5s5d)$~$^3$D$_1$ &3405 & 2.377 & & & & \\ 
$(5s5p)$~$^3$P$^o_0$ $\leftrightarrow$ $(5s5d)$~$^3$D$_2$ &3403& & & & & 2.426 \\ 
$(5s5p)$~$^3$P$^o_0$ $\leftrightarrow$ $(5s5d)$~$^3$D$_3$ &3401& & & 1.041$\times$10$^{2}$ & & \\ 
$(5s5p)$~$^3$P$^o_0$ $\leftrightarrow$ $(5s6d)$~$^3$D$_1$ &2838& 1.108 & & & & \\ 
$(5s5p)$~$^3$P$^o_0$ $\leftrightarrow$ $(5s6d)$~$^3$D$_2$ &2837& & & & & 1.088 \\ 
$(5s5p)$~$^3$P$^o_0$ $\leftrightarrow$ $(5s6d)$~$^3$D$_3$ &2837& & & 2.490$\times$10$^{1}$ & & \\
$(5s5p)$~$^3$P$^o_1$ $\leftrightarrow$ $(5s5p)$~$^3$P$^o_2$ &8.540$\times$10$^{4}$& & 1.828$\times$10$^{1}$ & & 1.578 & \\ 
$(5s5p)$~$^3$P$^o_1$ $\leftrightarrow$ $(5s5p)$~$^1$P$^o_1$ &7671& & 1.476 & & 5.368$\times$10$^{-2}$ & \\ 
$(5s5p)$~$^3$P$^o_1$ $\leftrightarrow$ $(5s6s)$~$^3$S$_1$ &4801& 2.621 & & & & 5.588 \\ 
$(5s5p)$~$^3$P$^o_1$ $\leftrightarrow$ $(5s5d)$~$^1$D$_2$ &3501& 1.816$\times$10$^{-1}$ & & 4.494 & & 1.097$\times$10$^{1}$ \\ 
$(5s5p)$~$^3$P$^o_1$ $\leftrightarrow$ $(5s5d)$~$^3$D$_1$ &3469& 2.098 & & & & 2.077 \\ 
$(5s5p)$~$^3$P$^o_1$ $\leftrightarrow$ $(5s5d)$~$^3$D$_2$ &3467& 3.623 & & 1.075$\times$10$^{2}$ & & 9.772 \\ 
$(5s5p)$~$^3$P$^o_1$ $\leftrightarrow$ $(5s5d)$~$^3$D$_3$ &3465& & & 1.518$\times$10$^{2}$ & & 3.198 \\ 
$(5s5p)$~$^3$P$^o_1$ $\leftrightarrow$ $(5s6d)$~$^3$D$_1$ &2882& 9.627$\times$10$^{-1}$ & & & & 9.636$\times$10$^{-1}$ \\
$(5s5p)$~$^3$P$^o_1$ $\leftrightarrow$ $(5s6d)$~$^3$D$_2$ &2882& 1.677 & & 2.462$\times$10$^{1}$ & & 4.461 \\ 
$(5s5p)$~$^3$P$^o_1$ $\leftrightarrow$ $(5s6d)$~$^3$D$_3$ &2881& & & 3.509$\times$10$^{1}$ & & 1.474 \\
$(5s5p)$~$^3$P$^o_2$ $\leftrightarrow$ $(5s6s)$~$^3$S$_1$ &5087& 3.572 & & & & 1.075$\times$10$^{1}$ \\ 
$(5s5p)$~$^3$P$^o_2$ $\leftrightarrow$ $(5s5d)$~$^1$D$_2$ &3651& 9.339$\times$10$^{-2}$ & & 3.778 & & 1.030$\times$10$^{1}$ \\ 
$(5s5p)$~$^3$P$^o_2$ $\leftrightarrow$ $(5s5d)$~$^3$D$_1$ &3615& 5.617$\times$10$^{-1}$ & & 1.547$\times$10$^{2}$ & & 1.033$\times$10$^{-1}$ \\ 
$(5s5p)$~$^3$P$^o_2$ $\leftrightarrow$ $(5s5d)$~$^3$D$_2$ &3614& 2.176 & & 1.630$\times$10$^{2}$ & & 7.103 \\ 
$(5s5p)$~$^3$P$^o_2$ $\leftrightarrow$ $(5s5d)$~$^3$D$_3$ &3611& 5.160 & & 1.263$\times$10$^{2}$ & & 2.133$\times$10$^{1}$ \\ 
$(5s5p)$~$^3$P$^o_2$ $\leftrightarrow$ $(5s6d)$~$^3$D$_1$ &2983& 2.514$\times$10$^{-1}$ & & 3.246$\times$10$^{1}$ & & 5.619$\times$10$^{-2}$ \\ 
$(5s5p)$~$^3$P$^o_2$ $\leftrightarrow$ $(5s6d)$~$^3$D$_2$ &2982& 9.796$\times$10$^{-1}$ & & 3.443$\times$10$^{1}$ & & 1.049$\times$10$^{1}$ \\ 
$(5s5p)$~$^3$P$^o_2$ $\leftrightarrow$ $(5s6d)$~$^3$D$_3$ &2981& 2.339 & & 2.693$\times$10$^{1}$ & & 9.695 \\ 
$(5s5p)$~$^1$P$^o_1$ $\leftrightarrow$ $(5s5d)$~$^1$D$_2$ &6440& 5.754 & & 4.081$\times$10$^{2}$ & & 1.458$\times$10$^{1}$ \\  
$(5s5p)$~$^1$P$^o_1$ $\leftrightarrow$ $(5s6d)$~$^1$D$_2$ &4662& 9.915$\times$10$^{-1}$      & & 5.644$\times$10$^{1}$        &  & 4.100  \\
$(5s5p)$~$^1$P$^o_1$ $\leftrightarrow$ $(5s7d)$~$^1$D$_2$ &4140& 1.815$\times$10$^{-1}$      & & 4.067$\times$10$^{1}$        &  & 2.206  \\
$(5s5d)$~$^3$D$_1$ $\leftrightarrow$ $(5s5d)$~$^3$D$_2$ &8.333$\times$10$^{6}$& & 7.690$\times$10$^{1}$ & & 2.121 & \\ 
$(5s5d)$~$^3$D$_2$ $\leftrightarrow$ $(5s5d)$~$^3$D$_3$ &5.556$\times$10$^{6}$& & 8.251$\times$10$^{1}$ & & 2.159 & \\ 
$(5s7p)$~$^3$P$^o_2$ $\leftrightarrow$ $(5s6d)$~$^3$D$_3$ &3.663$\times$10$^{5}$& 3.364$\times$10$^{1}$ & & 8.777$\times$10$^{3}$ & & 1.376$\times$10$^{2}$ \\ 
$(5s7p)$~$^3$P$^o_2$ $\leftrightarrow$ $(5s5f)$~$^3$F$^o_4$ &3.333$\times$10$^{4}$& & 7.251$\times$10$^{2}$ & &  & \\ 
$(5s6d)$~$^3$D$_1$ $\leftrightarrow$ $(5s6d)$~$^3$D$_2$ &1.667$\times$10$^{7}$& & 3.248$\times$10$^{2}$ & & 2.121 & \\ 
$(5s6d)$~$^3$D$_2$ $\leftrightarrow$ $(5s6d)$~$^3$D$_3$ &1.250$\times$10$^{7}$& & 3.480$\times$10$^{2}$ & & 2.160 & \\ 
$(5s6d)$~$^3$D$_2$ $\leftrightarrow$ $(5s4f)$~$^3$F$^o_3$ &4.405$\times$10$^{5}$& 2.349$\times$10$^{1}$ & & 3.263$\times$10$^{3}$ & & 1.114$\times$10$^{2}$ \\ 
$(5s6d)$~$^3$D$_3$ $\leftrightarrow$ $(5s4f)$~$^3$F$^o_4$ &4.566$\times$10$^{5}$& 3.140$\times$10$^{1}$ & & 9.041$\times$10$^{3}$ & & 1.639$\times$10$^{2}$ \\
$(5s6d)$~$^3$D$_3$ $\leftrightarrow$ $(5s5f)$~$^3$F$^o_4$ &3.667$\times$10$^{4}$& 2.550$\times$10$^{1}$ & & 1.314$\times$10$^{4}$ & & 1.33$\times$10$^{2}$ \\ 
$(5s4f)$~$^3$F$^o_2$ $\leftrightarrow$ $(5s4f)$~$^3$F$^o_3$ &$>$10$^{8}$& & 1.968$\times$10$^{2}$ & & 2.311 & \\ 
$(5s4f)$~$^3$F$^o_3$ $\leftrightarrow$ $(5s4f)$~$^3$F$^o_4$ &$>$10$^{8}$& & 2.009$\times$10$^{2}$ & & 2.323 & \\ 
$(5s8s)$~$^3$S$_1$ $\leftrightarrow$ $(5s8p)$~$^3$P$^o_2$ &8.382$\times$10$^{4}$& 3.003$\times$10$^{1}$ & & 9.787$\times$10$^{-1}$& & 9.009$\times$10$^{1}$ \\  
$(5s8s)$~$^3$S$_1$ $\leftrightarrow$ $(5s7d)$~$^3$D$_2$ &7.628$\times$10$^{4}$& & 8.190$\times$10$^{2}$ & & & \\ 
$(5s8s)$~$^3$S$_1$ $\leftrightarrow$ $(5s7d)$~$^3$D$_3$ &7.605$\times$10$^{4}$& & 9.685$\times$10$^{2}$ & & & \\ 
$(5s8s)$~$^3$S$_1$ $\leftrightarrow$ $(5s5f)$~$^3$F$^o_4$ &7.082$\times$10$^{4}$& & & 3.117$\times$10$^{4}$ & & \\ 
$(5s8s)$~$^1$S$_0$ $\leftrightarrow$ $(5s7d)$~$^1$D$_2$ &1.072$\times$10$^{5}$& & 8.735$\times$10$^{2}$ & & & \\
$(5s8p)$~$^3$P$^o_1$ $\leftrightarrow$ $(5s8p)$~$^3$P$^o_2$ &3.030$\times$10$^{6}$& & 9.506$\times$10$^{2}$ & & 1.572 & \\ 
\end{tabular}
\end{ruledtabular}
\end{table*}

\begin{table*}[t]
\centering
\rowcolors{2}{gray!10}{white}
\begin{ruledtabular}
\begin{tabular}{ccccccc}
States & $\lambda$ ($\SI{}{\angstrom}$) & E1       & E2 & E3 & M1 & M2 \\ \hline \addlinespace
$(5s8p)$~$^3$P$^o_1$ $\leftrightarrow$ $(5s7d)$~$^3$D$_2$ &6.622$\times$10$^{5}$& 3.523$\times$10$^{1}$ & & 2.246$\times$10$^{4}$ & & 8.941$\times$10$^{1}$ \\ 
$(5s8p)$~$^3$P$^o_1$ $\leftrightarrow$ $(5s7d)$~$^3$D$_3$ &6.452$\times$10$^{5}$& & & 3.177$\times$10$^{4}$ & & 3.973$\times$10$^{1}$ \\ 
$(5s8p)$~$^3$P$^o_2$ $\leftrightarrow$ $(5s7d)$~$^3$D$_1$ &8.696$\times$10$^{5}$& 5.260 & & 3.031$\times$10$^{4}$ & & 4.645$\times$10$^{-2}$ \\ 
$(5s8p)$~$^3$P$^o_2$ $\leftrightarrow$ $(5s7d)$~$^3$D$_2$ &8.475$\times$10$^{5}$& 2.039$\times$10$^{1}$ & & 3.197$\times$10$^{4}$ & & 6.282$\times$10$^{1}$ \\ 
$(5s8p)$~$^3$P$^o_2$ $\leftrightarrow$ $(5s7d)$~$^3$D$_3$ &8.197$\times$10$^{5}$& 4.830$\times$10$^{1}$ & & 2.479$\times$10$^{4}$ & & 1.973$\times$10$^{2}$ \\ 
$(5s8p)$~$^3$P$^o_2$ $\leftrightarrow$ $(5s5f)$~$^3$F$^o_4$ &4.566$\times$10$^{5}$& & 1.305$\times$10$^{3}$ & & & \\ 
$(5s7d)$~$^1$D$_2$ $\leftrightarrow$ $(5s8p)$~$^1$P$^o_1$ &4.525$\times$10$^{5}$& 3.696$\times$10$^{1}$ & & 3.712$\times$10$^{4}$ & & 8.151$\times$10$^{1}$ \\ 
$(5s7d)$~$^1$D$_2$ $\leftrightarrow$ $(5s5f)$~$^3$F$^o_3$ &3.906$\times$10$^{5}$& 3.121$\times$10$^{1}$ & & 2.381$\times$10$^{4}$ & & 1.487$\times$10$^{2}$ \\ 
$(5s7d)$~$^1$D$_2$ $\leftrightarrow$ $(5s5f)$~$^3$F$^o_4$ &3.906$\times$10$^{5}$& & & 1.296$\times$10$^{2}$ & & 1.274$\times$10$^{2}$ \\ 
$(5s7d)$~$^3$D$_1$ $\leftrightarrow$ $(5s7d)$~$^3$D$_2$ &3.333$\times$10$^{7}$& & 6.784$\times$10$^{2}$ & & 2.121 & \\ 
$(5s7d)$~$^3$D$_1$ $\leftrightarrow$ $(5s5f)$~$^3$F$^o_2$ &9.615$\times$10$^{5}$& 3.495$\times$10$^{1}$ & & 1.462$\times$10$^{4}$ & & 5.829$\times$10$^{1}$ \\ 
$(5s7d)$~$^3$D$_2$ $\leftrightarrow$ $(5s7d)$~$^3$D$_3$ &2.500$\times$10$^{7}$& & 7.257$\times$10$^{2}$ & & 2.160 & \\ 
$(5s7d)$~$^3$D$_2$ $\leftrightarrow$ $(5s5f)$~$^3$F$^o_3$ &9.901$\times$10$^{5}$& 3.110$\times$10$^{1}$ & & 9.216$\times$10$^{3}$ & & 1.764$\times$10$^{2}$ \\ 
$(5s7d)$~$^3$D$_2$ $\leftrightarrow$ $(5s9s)$~$^3$S$_1$ &1.451$\times$10$^{5}$& & 7.316$\times$10$^{2}$ & & & \\ 
$(5s7d)$~$^3$D$_3$ $\leftrightarrow$ $(5s8p)$~$^1$P$^o_1$ &1.613$\times$10$^{6}$& & & 3.248$\times$10$^{3}$ & & 1.103$\times$10$^{2}$ \\ 
$(5s7d)$~$^3$D$_3$ $\leftrightarrow$ $(5s5f)$~$^3$F$^o_4$ &1.031$\times$10$^{6}$& 5.110$\times$10$^{1}$ & & 3.132$\times$10$^{4}$ & & 2.666$\times$10$^{2}$ \\ 
$(5s7d)$~$^3$D$_3$ $\leftrightarrow$ $(5s9s)$~$^3$S$_1$ &1.460$\times$10$^{5}$& & 8.682$\times$10$^{2}$ & & & \\ 
$(5s8p)$~$^1$P$^o_1$ $\leftrightarrow$ $(5s5f)$~$^3$F$^o_3$ &2.857$\times$10$^{6}$& & 8.095$\times$10$^{2}$ & & & \\ 
$(5s5f)$~$^3$F$^o_2$ $\leftrightarrow$ $(5s5f)$~$^3$F$^o_3$ &$>$10$^{8}$& & 4.029$\times$10$^{2}$ & & 1.878 & \\ 
$(5s5f)$~$^3$F$^o_3$ $\leftrightarrow$ $(5s5f)$~$^3$F$^o_4$ &$>$10$^{8}$& & 4.112$\times$10$^{2}$ & & 1.890 & \\ 
$(5s5f)$~$^3$F$^o_4$ $\leftrightarrow$ $(5s9s)$~$^3$S$_1$ &1.701$\times$10$^{5}$& && 2.633$\times$10$^{4}$ & & \\ 

\end{tabular}
\end{ruledtabular}

\end{table*}

\subsection{Multipole Matrix Elements}

Table \ref{tab:mat_elem} presents the E1-E3 and M1-M2 transition matrix elements expressed in atomic units (a.u.), calculated using CI+MBPT. Empty entries indicate either that the transition is forbidden by selection rules or that the corresponding matrix element is smaller than $10^{-3}$~a.u. Coupling to the nuclear spin, as well as mixed multipole contributions (e.g., E1M1), are not included; consequently, the clock transitions of fermionic isotopes are absent in this table.

The code generates on the order of $10^4$ transitions. In the following, we focus on transitions most relevant to quantum gases, metrology, and atom trapping, as outlined in the Introduction. We highlight (i) transitions involving the ground state $(5s^2) \, ^1$S$_0$, (ii) the ten strongest transitions of each multipole type, and (iii) we make sure all the transitions available on the NIST database are present. Transitions for which all five multipole matrix elements vanish are omitted. AMBiT requires the simulation parameters to be compiled in an \texttt{.input} file, which is provided as Supplementary Material. This file enables readers to reproduce the calculations and examine additional transitions not reported in the main text. To facilitate the identification of individual transitions, their corresponding wavelengths, expressed in ångströms, are provided. These wavelength values are taken from the NIST database~\cite{Burns1956,Brown1975,VidolovaAngelova1996,Moore1971,NIST_ASD}.

In general, when a matrix element exists, its magnitude tends to increase for transitions involving higher shells. Similarly, when the interaction is expressed in atomic units, whether electric or magnetic, higher multipole orders correspond to larger matrix elements.

The $(5s^2)$~$^1$S$_0$ $\leftrightarrow$ $(5s5p)$~$^1$P$^o_1$ transition exhibits a relatively large E1 matrix element; this is the broad cooling transition of neutral cadmium at 229\,nm. However, the transition linewidth must account not only for the intrinsic width but also for the wavelength, as discussed in the following section. The second cooling transition is also present, albeit with a smaller amplitude. Among the clock transitions, the $^3$P$_0$ state has vanishing matrix elements for all multipolar orders, whereas the $(5s5p)$~$^3$P$^o_2$ state exhibits a nonzero M2 matrix element, as detailed in the next section.

\section{Radiative decay rates}
\subsection{Theory}
The total decay rate is the sum of all multipole contributions. The type and number of multipoles that contribute to the sum is limited by selection rules. Therefore, for a transition from an initial state $\ket{a}$ of higher energy to a final state $\ket{b}$ of lower energy, the Einstein $A$-coefficient $A_{ba}$ is the sum of each multipole element~\cite{johnson2007atomic}
\begin{equation}
\label{eq11}
    A_{ab} = 8\pi\, \alpha\,\omega\sum_ {\eta , \xi}\left|\int \psi_b^{\dagger}\, \boldsymbol{\alpha}\cdot \boldsymbol{a}_{\eta}^{(\xi)}\psi_a\,d^3r \right|^2 
\end{equation}

with $\alpha$ the fine structure constant, $\omega$ the light field frequency, $\psi_a$ and $\psi_b$ the Dirac wavefunctions of respectively the states $a$ and $b$, $\xi = E$ to designate the electric multipole potentials of order $\eta$ of the vector potential $\boldsymbol{a}_{\eta}^{(E)}$ and $\xi = M$ to designate the magnetic multipole poential of order $\eta$ of the vector potential $\boldsymbol{a}_{\eta}^{(M)}$.

The lifetime $\tau$ in a state $a$ is then the sum over all the possible transitions from this state
\begin{equation}
\label{eq12}
    \tau = \sum_{b\leq a} \frac{1}{A_{ba}}
\end{equation}

Using equations \ref{eq9} and \ref{eq10}, the decay rates of each multipole channel $A_{ba}^{(\xi\eta)}$ can be expressed either in SI units, or in customary units, where the multipole matrix elements are expressed in a.u. and the wavelength in ångström~\cite{johnson2007atomic,Drake1996}

\begin{table}[htbp!]
\caption{\justifying Comparison of the general formulas for multipole transition probabilities $A_{ba}^{(E\eta)}$ and $A_{ba}^{(M\eta)}$ expressed in SI units and customary atomic units.}
\label{tab:MultipoleA}
\begin{ruledtabular}
\begin{tabular}{c|c|c}
                   & SI units & Customary units \\ \hline

\makecell{Electric \\ Multipole} & 
$A_{ba}^{(E\eta)} = \mathscr{C}_{\eta}\dfrac{S^{(E\eta)}}{2h\epsilon_0\lambda^{2\eta+1}[J]}$ & 
$A_{ba}^{(E\eta)} = \mathscr{N}^{(E)}_{\eta}\dfrac{S^{(E\eta)}}{\lambda^{2\eta+1}[J]}$ \\ \hline

\makecell{Magnetic \\ Multipole} & 
$A_{ba}^{(M\eta)} = \mathscr{C}_{\eta}\dfrac{\mu_0 S^{(M\eta)}}{2h\lambda^{2\eta+1}[J]}$ &
$A_{ba}^{(M\eta)} = \mathscr{N}^{(M)}_{\eta}\dfrac{S^{(M\eta)}}{\lambda^{2\eta+1}[J]}$ \\
\end{tabular}
\end{ruledtabular}
\end{table}

In table \ref{tab:MultipoleA}, $\left[J\right]=2J+1$ denotes the statistical weights with $J$ being the total angular momentum of the upper level. $e$ is the unit of electric charge, $a_0$ the Bohr radius, $h$ the Planck constant and $\epsilon_0$ the vacuum permittivity. $\mu_0$ and $\mu_B$ are respectively the vacuum permeability and the Bohr magneton.

The line strength of the transition $a \leftrightarrow b $ is $S^{(\xi\eta)}~=~\left|\bra{b}\boldsymbol{O}^{(\xi\eta)}\ket{a}\right|^2 $. $\mathscr{C}_{\eta}$ and $\mathscr{N}^{(\xi)}_{\eta}$, the two constants of order $\eta$ used to adapt the decay rates from one unit system to the other, are shown in table \ref{tab:constants_eta}.

\begin{table}[htbp!]
\caption{\justifying Constants of order $\eta$ involved in the calculations of the multipole channels.}
\label{tab:constants_eta}
\resizebox{0.51\textwidth}{!}{%
\begin{tabular}{c|c|c}
\hline \hline
$\mathscr{C}_\eta$ &
$\mathscr{N}^{(E)}_\eta$ &
$\mathscr{N}^{(M)}_\eta$ \\ \hline
$\dfrac{(2\eta+2)(2\eta+1)(2\pi)^{2\eta+1}}{\eta!\,(2\eta+1)!!^2}$ &
$\mathscr{C}_\eta \dfrac{e^2 a_0^{2\eta}}{2 h \epsilon_0}(10^{20\eta+10})$ &
$\mathscr{C}_\eta \dfrac{\mu_0 \mu_B^2 a_0^{2\eta-2}}{2h}(10^{20\eta+10})$ \\ \hline \hline
\end{tabular}%
}
\end{table}

\subsection{Results}

\begin{table*}[hbtp!]
\centering
\rowcolors{2}{gray!10}{white}
\begin{ruledtabular}
\caption{\justifying Spontaneous decay rate $A_{ba}$ in s$^{-1}$ expressed as the sum of the first five multipole decay rates $A_{ba}^{(\xi\eta)}$ and compared to the NIST database for the strongest lines.}
\label{tab:decay_rate}
\begin{tabular}{cccccccc} 
States & $A_{ba}^{(E1)}$ & $A_{ba}^{(E2)}$ & $A_{ba}^{(E3)}$ & $A_{ba}^{(M1)}$ & $A_{ba}^{(M2)}$ & $A_{ba}$ & NIST   \\ \hline \addlinespace
$(5s^2)$~$^1$S$_0$ $\leftrightarrow$ $(5s5p)$~$^3$P$^o_1$   & 6.192$\times$10$^{5}$ & & & &&6.192$\times$10$^{5}$&4.06(102)$\times$10$^{5}$ \\ 
$(5s^2)$~$^1$S$_0$ $\leftrightarrow$ $(5s5p)$~$^3$P$^o_2$   & & & & & 2.183$\times$10$^{-3}$ &2.183$\times$10$^{-3}$&\\
$(5s^2)$~$^1$S$_0$ $\leftrightarrow$ $(5s5p)$~$^1$P$^o_1$   & 8.079$\times$10$^{8}$ & & & & &8.079$\times$10$^{8}$&5.3(13)$\times$10$^{8}$\\ 
$(5s^2)$~$^1$S$_0$ $\leftrightarrow$ $(5s5d)$~$^1$D$_2$     & &2.350$\times$10$^{3}$ & & &&2.350$\times$10$^{3}$&\\ 
$(5s^2)$~$^1$S$_0$ $\leftrightarrow$ $(5s5d)$~$^3$D$_2$     & &1.129 & & &&1.129&\\ 
$(5s^2)$~$^1$S$_0$ $\leftrightarrow$ $(5s6p)$~$^1$P$^o_1$   & 1.202$\times$10$^{8}$ & & & & &1.202$\times$10$^{8}$&\\ 
$(5s^2)$~$^1$S$_0$ $\leftrightarrow$ $(5s6d)$~$^1$D$_2$     & & 1.011$\times$10$^{3}$ & & & &1.011$\times$10$^{3}$&\\ 
$(5s^2)$~$^1$S$_0$ $\leftrightarrow$ $(5s6d)$~$^3$D$_2$     & & 7.235$\times$10$^{-2}$ & & & &7.235$\times$10$^{-2}$&\\
$(5s^2)$~$^1$S$_0$ $\leftrightarrow$ $(5s7p)$~$^1$P$^o_1$   & 4.089$\times$10$^{7}$ & & & & &4.089$\times$10$^{7}$&\\ 
$(5s^2)$~$^1$S$_0$ $\leftrightarrow$ $(5s8p)$~$^1$P$^o_1$   & 2.704$\times$10$^{7}$ & & & & &2.704$\times$10$^{7}$&\\ 
$(5s5p)$~$^3$P$^o_0$ $\leftrightarrow$ $(5s5p)$~$^3$P$^o_1$ & & & & 2.854$\times$10$^{-3}$ & &2.854$\times$10$^{-3}$&\\
$(5s5p)$~$^3$P$^o_0$ $\leftrightarrow$ $(5s5p)$~$^3$P$^o_2$ & & 4.796$\times$10$^{-5}$ & & & &4.796$\times$10$^{-5}$&\\ 
$(5s5p)$~$^3$P$^o_0$ $\leftrightarrow$ $(5s5p)$~$^1$P$^o_1$ & & & & 8.526$\times$10$^{-2}$ & &8.526$\times$10$^{-2}$&\\ 
$(5s5p)$~$^3$P$^o_0$ $\leftrightarrow$ $(5s6s)$~$^3$S$_1$   & 1.442$\times$10$^{7}$ & & & & &1.442$\times$10$^{7}$&1.3(3)$\times$10$^{7}$\\ 
$(5s5p)$~$^3$P$^o_0$ $\leftrightarrow$ $(5s5d)$~$^1$D$_2$   & & & & & 3.579$\times$10$^{-4}$ &3.579$\times$10$^{-4}$&\\ 
$(5s5p)$~$^3$P$^o_0$ $\leftrightarrow$ $(5s5d)$~$^3$D$_1$   & 9.665$\times$10$^{7}$ & & & & &9.665$\times$10$^{7}$&7.7(19)$\times$10$^{7}$\\ 
$(5s5p)$~$^3$P$^o_0$ $\leftrightarrow$ $(5s5d)$~$^3$D$_2$   & & & & & 3.846$\times$10$^{-5}$ &3.846$\times$10$^{-5}$&\\ 
$(5s5p)$~$^3$P$^o_0$ $\leftrightarrow$ $(5s5d)$~$^3$D$_3$   & & & 9.250$\times$10$^{-13}$ & & &9.250$\times$10$^{-13}$&\\ 
$(5s5p)$~$^3$P$^o_0$ $\leftrightarrow$ $(5s6d)$~$^3$D$_1$   & 3.627$\times$10$^{7}$ & & & & &3.627$\times$10$^{7}$&2.8(14)$\times$10$^{7}$\\ 
$(5s5p)$~$^3$P$^o_0$ $\leftrightarrow$ $(5s6d)$~$^3$D$_2$   & & & & & 1.921$\times$10$^{-5}$ &1.921$\times$10$^{-5}$&\\ 
$(5s5p)$~$^3$P$^o_0$ $\leftrightarrow$ $(5s6d)$~$^3D_3$   & & & 1.883$\times$10$^{-5}$ & & &1.883$\times$10$^{-5}$&\\
$(5s5p)$~$^3$P$^o_1$ $\leftrightarrow$ $(5s5p)$~$^3$P$^o_2$ & & 1.648$\times$10$^{-5}$ & & 2.157$\times$10$^{-2}$ &&2.159$\times$10$^{-2}$& \\ 
$(5s5p)$~$^3$P$^o_1$ $\leftrightarrow$ $(5s5p)$~$^1$P$^o_1$ & & 3.062$\times$10$^{-2}$ & & 5.740$\times$10$^{-2}$ & &8.802$\times$10$^{-2}$&\\ 
$(5s5p)$~$^3$P$^o_1$ $\leftrightarrow$ $(5s6s)$~$^3$S$_1$   & 4.193$\times$10$^{7}$ & & & & 6.085$\times$10$^{-5}$ &4.192$\times$10$^{7}$&4.1(10)$\times$10$^{7}$\\ 
$(5s5p)$~$^3$P$^o_1$ $\leftrightarrow$ $(5s5d)$~$^1$D$_2$   & 3.114$\times$10$^{5}$ & & 1.970$\times$10$^{-7}$ & & 6.823$\times$10$^{-4}$ &3.114$\times$10$^{5}$&\\ 
$(5s5p)$~$^3$P$^o_1$ $\leftrightarrow$ $(5s5d)$~$^3$D$_1$   & 7.120$\times$10$^{7}$ & & & & 4.268$\times$10$^{-5}$ &7.120$\times$10$^{7}$&6.7(34)$\times$10$^{7}$\\ 
$(5s5p)$~$^3$P$^o_1$ $\leftrightarrow$ $(5s5d)$~$^3D_2$   & 1.276$\times$10$^{8}$ & & 1.207$\times$10$^{-4}$ & & 5.685$\times$10$^{-4}$ &1.276$\times$10$^{8}$&1.2(6)$\times$10$^{8}$\\ 
$(5s5p)$~$^3$P$^o_1$ $\leftrightarrow$ $(5s5d)$~$^3$D$_3$   & & & 1.719$\times$10$^{-4}$ & & 4.362$\times$10$^{-5}$ &2.156$\times$10$^{-4}$&\\ 
$(5s5p)$~$^3$P$^o_1$ $\leftrightarrow$ $(5s6d)$~$^3$D$_1$   & 2.615$\times$10$^{7}$ & & & & 2.321$\times$10$^{-5}$ &2.615$\times$10$^{7}$&2.4(12)$\times$10$^{7}$\\
$(5s5p)$~$^3$P$^o_1$ $\leftrightarrow$ $(5s6d)$~$^3$D$_2$   & 4.760$\times$10$^{7}$ & & 2.308$\times$10$^{-5}$ & & 2.985$\times$10$^{-4}$ &4.760$\times$10$^{7}$&4.2(21)$\times$10$^{7}$\\ 
$(5s5p)$~$^3$P$^o_1$ $\leftrightarrow$ $(5s6d)$~$^3$D$_3$   & & & 3.358$\times$10$^{-5}$ & & 2.332$\times$10$^{-5}$ &5.690$\times$10$^{-5}$& \\
$(5s5p)$~$^3$P$^o_2$ $\leftrightarrow$ $(5s6s)$~$^3$S$_1$   & 6.545$\times$10$^{7}$ & & & & 1.686$\times$10$^{-4}$ &6.545$\times$10$^{7}$&5.6(14)$\times$10$^{7}$\\ 
$(5s5p)$~$^3$P$^o_2$ $\leftrightarrow$ $(5s5d)$~$^1$D$_2$   &7.261$\times$10$^{4}$ & & 1.038$\times$10$^{-7}$ & & 4.877$\times$10$^{-4}$ &7.261$\times$10$^{4}$& \\ 
$(5s5p)$~$^3$P$^o_2$ $\leftrightarrow$ $(5s5d)$~$^3$D$_1$   & 4.510$\times$10$^{6}$ & & 3.109$\times$10$^{-4}$ & & 8.591$\times$10$^{-8}$ &4.510$\times$10$^{6}$& \\ 
$(5s5p)$~$^3$P$^o_2$ $\leftrightarrow$ $(5s5d)$~$^3$D$_2$   & 4.064$\times$10$^{7}$ & & 2.075$\times$10$^{-4}$ & & 2.441$\times$10$^{-4}$ &4.064$\times$10$^{7}$&3.5(18)$\times$10$^{7}$\\ 
$(5s5p)$~$^3$P$^o_2$ $\leftrightarrow$ $(5s5d)$~$^3$D$_3$   & 1.637$\times$10$^{8}$ & & 8.951$\times$10$^{-5}$ & & 1.579$\times$10$^{-3}$ &1.637$\times$10$^{8}$&1.3(7)$\times$10$^{8}$\\
$(5s5p)$~$^3$P$^o_2$ $\leftrightarrow$ $(5s6d)$~$^3$D$_1$   & 1.608$\times$10$^{6}$ & & 5.255$\times$10$^{-5}$ & & 6.644$\times$10$^{-8}$ &1.608$\times$10$^{6}$&\\ 
$(5s5p)$~$^3$P$^o_2$ $\leftrightarrow$ $(5s6d)$~$^3$D$_2$   & 1.466$\times$10$^{7}$ & & 3.556$\times$10$^{-5}$ & & 1.392$\times$10$^{-3}$ &1.466$\times$10$^{7}$&1.5(8)$\times$10$^{7}$\\ 
$(5s5p)$~$^3$P$^o_2$ $\leftrightarrow$ $(5s6d)$~$^3$D$_3$   & 5.977$\times$10$^{7}$ & & 1.557$\times$10$^{-5}$ & & 8.506$\times$10$^{-4}$ &5.977$\times$10$^{7}$&5.9(29)$\times$10$^{7}$\\ 
$(5s5p)$~$^1$P$^o_1$ $\leftrightarrow$ $(5s5d)$~$^1$D$_2$   & 5.023$\times$10$^{7}$ & & 2.280$\times$10$^{-5}$ & & 5.723$\times$10$^{-5}$ &5.023$\times$10$^{7}$&5.9(29)$\times$10$^{7}$\\  
$(5s5p)$~$^1$P$^o_1$ $\leftrightarrow$ $(5s6d)$~$^1$D$_2$   & 3.931$\times$10$^{6}$ & & 4.186$\times$10$^{-6}$ & & 2.276$\times$10$^{-5}$ &3.932$\times$10$^{6}$ &5.5(14)$\times$10$^{6}$ \\   
$(5s5p)$~$^1$P$^o_1$ $\leftrightarrow$ $(5s7d)$~$^1$D$_2$   & 1.881$\times$10$^{5}$ & & 4.991$\times$10$^{-6}$ & & 1.193$\times$10$^{-5}$ &1.885$\times$10$^{5}$  &4.7(24)$\times$10$^{6}$ \\
($(5s5p)$~$^1$P$^o_1$ $\leftrightarrow$ $(5s7d)$~$^1$D$_2$) &(1.035$\times$10$^{6}$)& &(4.991$\times$10$^{-6}$)& &(1.193$\times$10$^{-5}$)&(1.035$\times$10$^{6}$)  &4.7(24)$\times$10$^{6}$ \\
$(5s5d)$~$^3$D$_1$ $\leftrightarrow$ $(5s5d)$~$^3$D$_2$     & & 3.300$\times$10$^{-14}$ & & 4.194$\times$10$^{-8}$ & &4.194$\times$10$^{-8}$&\\ 
$(5s5d)$~$^3$D$_2$ $\leftrightarrow$ $(5s5d)$~$^3$D$_3$     & & 2.057$\times$10$^{-13}$ & & 1.047$\times$10$^{-7}$ & &1.047$\times$10$^{-7}$&\\ 
$(5s7p)$~$^3$P$^o_2$ $\leftrightarrow$ $(5s6d)$~$^3$D$_3$   & 6.664$\times$10$^{3}$ & & 3.911$\times$10$^{-15}$ & & 6.116$\times$10$^{-12}$ &6.664$\times$10$^{3}$&\\ 
$(5s7p)$~$^3$P$^o_2$ $\leftrightarrow$ $(5s5f)$~$^3$F$^o_4$ & & 1.591 & &  & &1.591&\\ 
$(5s6d)$~$^3$D$_1$ $\leftrightarrow$ $(5s6d)$~$^3$D$_2$     & & 1.836$\times$10$^{-14}$ & & 5.239$\times$10$^{-9}$ & &5.239$\times$10$^{-9}$&\\ 
$(5s6d)$~$^3$D$_2$ $\leftrightarrow$ $(5s6d)$~$^3$D$_3$     & & 6.349$\times$10$^{-14}$ & & 9.205$\times$10$^{-9}$ & &9.205$\times$10$^{-9}$&\\ 
$(5s6d)$~$^3$D$_2$ $\leftrightarrow$ $(5s4f)$~$^3$F$^o_3$   & 1.868$\times$10$^{3}$ & &1.486$\times$10$^{-16}$ & & 1.594$\times$10$^{-12}$ &1.868$\times$10$^{3}$&\\ 
$(5s6d)$~$^3$D$_3$ $\leftrightarrow$ $(5s4f)$~$^3$F$^o_4$   & 2.331$\times$10$^{3}$ & & 6.903$\times$10$^{-16}$ & & 2.243$\times$10$^{-12}$ &2.331$\times$10$^{3}$&\\
$(5s6d)$~$^3$D$_3$ $\leftrightarrow$ $(5s5f)$~$^3$F$^o_4$   & 2.968$\times$10$^{6}$ & & 6.766$\times$10$^{-8}$ & & 4.420$\times$10$^{-7}$ &2.968$\times$10$^{6}$&\\ 
$(5s4f)$~$^3$F$^o_2$ $\leftrightarrow$ $(5s4f)$~$^3$F$^o_3$ & & 0 & & 0 &&0& \\ 
$(5s4f)$~$^3$F$^o_3$ $\leftrightarrow$ $(5s4f)$~$^3$F$^o_4$ & & 0 & & 0 &&0& \\ 
$(5s8s)$~$^3$S$_1$ $\leftrightarrow$ $(5s8p)$~$^3$P$^o_2$   & 6.205$\times$10$^{5}$ & & 2.072$\times$10$^{-18}$ & & 5.850$\times$10$^{-9}$ &6.205$\times$10$^{5}$&\\  
$(5s8s)$~$^3$S$_1$ $\leftrightarrow$ $(5s7d)$~$^3$D$_2$     & & 5.817$\times$10$^{-2}$ & & & &5.817$\times$10$^{-2}$&\\
$(5s8s)$~$^3$S$_1$ $\leftrightarrow$ $(5s7d)$~$^3$D$_3$     & & 5.899$\times$10$^{-2}$ & & &&5.899$\times$10$^{-2}$& \\
$(5s8s)$~$^3$S$_1$ $\leftrightarrow$ $(5s5f)$~$^3$F$^o_4$   & & & 3.799$\times$10$^{-9}$ & & &3.799$\times$10$^{-9}$&\\ 
$(5s8s)$~$^1$S$_0$ $\leftrightarrow$ $(5s7d)$~$^1$D$_2$     & & 1.207$\times$10$^{-2}$ & & & &1.207$\times$10$^{-2}$&\\
$(5s8p)$~$^3$P$^o_1$ $\leftrightarrow$ $(5s8p)$~$^3$P$^o_2$ & & 7.925$\times$10$^{-10}$ & & 4.792$\times$10$^{-7}$ &&4.800$\times$10$^{-7}$& \\ 
\end{tabular}
\end{ruledtabular}
\end{table*}

\begin{table*}[t]
\centering
\rowcolors{2}{gray!10}{white}
\begin{ruledtabular}
\begin{tabular}{cccccccc} 
States & $A_{ba}^{(E1)}$ & $A_{ba}^{(E2)}$ & $A_{ba}^{(E3)}$ & $A_{ba}^{(M1)}$ & $A_{ba}^{(M2)}$ & $A_{ba}$ & NIST   \\ \hline \addlinespace
$(5s8p)$~$^3$P$^o_1$ $\leftrightarrow$ $(5s7d)$~$^3$D$_2$   & 1.732$\times$10$^{3}$ & & 5.682$\times$10$^{-16}$ & & 1.872$\times$10$^{-13}$ &1.732$\times$10$^{3}$& \\ 
$(5s8p)$~$^3$P$^o_1$ $\leftrightarrow$ $(5s7d)$~$^3$D$_3$   & & & 9.742$\times$10$^{-16}$ & & 3.007$\times$10$^{-14}$ &3.105$\times$10$^{-14}$& \\ 
$(5s8p)$~$^3$P$^o_2$ $\leftrightarrow$ $(5s7d)$~$^3$D$_1$   & 2.841$\times$10$^{1}$ & & 2.561$\times$10$^{-16}$ & & 2.157$\times$10$^{-20}$ &2.841$\times$10$^{1}$&\\ 
$(5s8p)$~$^3$P$^o_2$ $\leftrightarrow$ $(5s7d)$~$^3$D$_2$   & 2.767$\times$10$^{2}$ & & 2.047$\times$10$^{-16}$ & & 2.692$\times$10$^{-14}$ &2.767$\times$10$^{2}$&\\ 
$(5s8p)$~$^3$P$^o_2$ $\leftrightarrow$ $(5s7d)$~$^3$D$_3$   & 1.226$\times$10$^{3}$ & & 1.110$\times$10$^{-16}$ & & 2.241$\times$10$^{-13}$ &1.226$\times$10$^{3}$&\\ 
$(5s8p)$~$^3$P$^o_2$ $\leftrightarrow$ $(5s5f)$~$^3$F$^o_4$ & & 1.068$\times$10$^{-5}$ & & & &1.068$\times$10$^{-5}$&\\ 
$(5s7d)$~$^1$D$_2$ $\leftrightarrow$ $(5s8p)$~$^1$P$^o_1$   & 9.957$\times$10$^{3}$ & & 3.718$\times$10$^{-14}$ & & 1.741$\times$10$^{-12}$ &9.957$\times$10$^{3}$&\\ 
$(5s7d)$~$^1$D$_2$ $\leftrightarrow$ $(5s5f)$~$^3$F$^o_3$   & 4.731$\times$10$^{3}$ & & 1.836$\times$10$^{-14}$ & & 5.181$\times$10$^{-12}$ &4.731$\times$10$^{3}$&\\ 
$(5s7d)$~$^1$D$_2$ $\leftrightarrow$ $(5s5f)$~$^3$F$^o_4$   & & & 4.231$\times$10$^{-19}$ & & 2.958$\times$10$^{-12}$ &2.958$\times$10$^{-12}$&\\ 
$(5s7d)$~$^3$D$_1$ $\leftrightarrow$ $(5s7d)$~$^3$D$_2$     & & 2.506$\times$10$^{-15}$ & & 6.555$\times$10$^{-10}$ & &6.555$\times$10$^{-10}$&\\ 
$(5s7d)$~$^3$D$_1$ $\leftrightarrow$ $(5s5f)$~$^3$F$^o_2$   & 5.568$\times$10$^{2}$ & & 1.769$\times$10$^{-17}$ & & 1.233$\times$10$^{-14}$ &5.568$\times$10$^{2}$&\\ 
$(5s7d)$~$^3$D$_2$ $\leftrightarrow$ $(5s7d)$~$^3$D$_3$     & & 8.628$\times$10$^{-15}$ & & 1.151$\times$10$^{-9}$ &&1.151$\times$10$^{-9}$& \\ 
$(5s7d)$~$^3$D$_2$ $\leftrightarrow$ $(5s5f)$~$^3$F$^o_3$   & 2.884$\times$10$^{2}$ & & 4.091$\times$10$^{-18}$ & & 6.967$\times$10$^{-14}$ &2.884$\times$10$^{2}$&\\ 
$(5s7d)$~$^3$D$_2$ $\leftrightarrow$ $(5s9s)$~$^3$S$_1$     & & 3.107$\times$10$^{-3}$ & & & &3.107$\times$10$^{-3}$&\\
$(5s7d)$~$^3$D$_3$ $\leftrightarrow$ $(5s8p)$~$^1$P$^o_1$   & & & 3.893$\times$10$^{-20}$ & & 5.538$\times$10$^{-15}$ &5.538$\times$10$^{-15}$&\\ 
$(5s7d)$~$^3$D$_3$ $\leftrightarrow$ $(5s5f)$~$^3$F$^o_4$   & 5.363$\times$10$^{2}$ & & 2.768$\times$10$^{-17}$ & & 1.011$\times$10$^{-13}$ &5.363$\times$10$^{2}$&\\ 
$(5s7d)$~$^3$D$_3$ $\leftrightarrow$ $(5s9s)$~$^3$S$_1$     & & 4.242$\times$10$^{-3}$ & & & &4.242$\times$10$^{-3}$&\\
$(5s8p)$~$^1$P$^o_1$ $\leftrightarrow$ $(5s5f)$~$^3$F$^o_3$ & & 5.508$\times$10$^{-10}$ & & & &5.508$\times$10$^{-10}$&\\ 
$(5s5f)$~$^3$F$^o_2$ $\leftrightarrow$ $(5s5f)$~$^3$F$^o_3$ & & 0 & & 0 &&0& \\ 
$(5s5f)$~$^3$F$^o_3$ $\leftrightarrow$ $(5s5f)$~$^3$F$^o_4$ & & 0 & & 0 & &0&\\ 
$(5s5f)$~$^3$F$^o_4$ $\leftrightarrow$ $(5s9s)$~$^3$S$_1$   & && 1.764$\times$10$^{-11}$ & & &1.764$\times$10$^{-11}$&\\ 
\end{tabular}
\end{ruledtabular}
\end{table*}

Table \ref{tab:decay_rate} presents the decay rates (in s$^{-1}$) corresponding to each multipole contribution. These rates are obtained by applying the equations in Tables \ref{tab:MultipoleA} and \ref{tab:constants_eta} to the matrix elements listed in Table \ref{tab:mat_elem}, and are compared with the strongest available transition rates reported in the NIST database. An empty entry indicates that the corresponding matrix element is negligible. For some cases—particularly transitions within the $(5s4f)$~$^3$F$^o_J$ manifold—the energy separation between the levels is below 1\,cm$^{-1}$. Consequently, the corresponding wavelength is effectively infinite, yielding a negligible decay rate; this situation is denoted by “0” in Table \ref{tab:decay_rate}.

The decay rate is often denoted by $\Gamma$, although for experimental purposes the more relevant quantity is typically $A_{ba}^{(\xi\eta)}/2\pi$. Some strong transitions are absent from the NIST database, such as $(5s^2)$~$^1$S$_0 \leftrightarrow (5s7p)$~$^1$P$_1$ or $(5s^2)$~$^1S_0 \leftrightarrow (5s8p)$~$^1$P$_1$. This omission occurs because NIST does not report transitions below 200\,nm for cadmium, while this atom exhibits a substantial number of deep-UV transitions.

In most cases, a single multipole contribution dominates the decay, with contributions of the same order of magnitude being rare. However, exceptions do exist—for example, in the transition $(5s5p)$~$^3$P$^o_1 \leftrightarrow (5s6d)$~$^3$D$_3$, the E3 and M2 decay rates are of comparable strength.

It is noteworthy that some of the largest matrix elements correspond to decay rates well below the hertz level, highlighting the crucial role of the transition wavelength in determining the overall rate.

A comparison with the NIST database yields an average relative deviation of 18\,$\%$ and a maximum relative deviation of 53\,$\%$. Only one case differs by an order of magnitude: the $(5s5p)$~$^1$P$^o_1 \leftrightarrow (5s7d)$~$^1$D$_2$ transition. This discrepancy may stem from a misassignment in the spectroscopic interpretation of the AMBiT output. Because the computed states are strongly mixed, the identification relies on comparison of both the energy levels and the $g$-factors with the NIST reference. An alternative candidate assignment exists, located 2823\,cm$^{-1}$ from the reference (rather than 71\,cm$^{-1}$ as in our initial identification). For this second candidate, the calculated decay rate agrees with the NIST value. Both assignments are reported in Table \ref{tab:decay_rate}, with the alternative shown in brackets.

The experimental uncertainties reported in the 'NIST' column are taken from~\cite{Burns1956}. Accounting for a more precise determination of the line wavelength, for instance, using the NIST database, does not lead to results more accurate than those reported in Table \ref{tab:decay_rate}.

For physicists working with cadmium near the ground state $(5s^2)$~$^1$S$_0$, most relevant transitions are already well known. Nevertheless, we draw attention to the transition at 298.3\,nm with $\Gamma = 2\pi \times 256$\,kHz, which could prove useful for cooling or repumping schemes. Researchers from other fields may likewise find valuable reference data in this table for their applications.

\subsection{Transitions to metastable $^3$P$^o_0$ and $^3$P$^o_2$ states}
In atomic physics, a particular class of transitions, known as clock transitions, is of special interest for metrological applications. Such transitions must exhibit minimal sensitivity to external perturbations and involve two states with exceptionally long lifetimes. Cadmium, like other bosonic alkaline-earth-like elements, demonstrates extremely low sensitivity to stray electric and magnetic fields due to the absence of spin in its ground state~\cite{Ferrari2006}. Moreover, cadmium possesses two states with lifetimes sufficiently long to serve as clock transitions: $(5s5p)$~$^3$P$^o_0$ and $(5s5p)$~$^3$P$^o_2$.

For the fermionic isotope, the nonzero nuclear magnetic dipole moment induces hyperfine coupling, which mixes the $(5s5p)$~$^3$P$^o_0$ and $(5s5p)$~$^3$P$^o_2$ states with other states of the same parity, particularly the states $(5s5p)$~$^3$P$^o_1$ and $(5s5p)$~$^1$P$^o_1$. These mixed states can decay to the ground state via a standard electric dipole process, resulting in lifetimes on the order of 100\,s~\cite{Garstang1962}.

In contrast, this hyperfine-mediated decay mechanism is absent for bosonic isotopes, so the decay rate is dominated by alternative channels. The $(5s5p)$~$^3$P$^o_2$ state can decay to the $(5s^2)$~$^1$S$_0$ ground state via a magnetic quadrupole transition, with a rate on the order of the millihertz. This behavior is comparable to that observed in Mg, Ca, and Sr, which have similar electronic structures~\cite{Santra2004}. Such a transition has been extensively studied in Sr~\cite{Trautmann2023}. Since the $(5s5p)$~$^3$P$^o_1$ $\leftrightarrow$ $(5s5p)$~$^3$P$^o_2$ transition is estimated to decay roughly an order of magnitude faster, the $(5s5p)$~$^3$P$^o_2$ state may not be an ideal candidate for a clock transition.

On the other hand, the $(5s^2)$~$^1$S$_0$ $\leftrightarrow$ $(5s5p)$~$^3$P$^o_0$ transition is strictly forbidden under all single-photon multipole channels due to the $J=0 \not\to J=0$ selection rule. The only allowed decay channel is the two-photon E1M1 process, in which the $(5s5p)$~$^3$P$^o_0$ state is coupled virtually to an intermediate state of the same parity via an M1 transition, followed by an E1 transition to the ground state~\cite{Craig1998MQED}.

\begin{equation}
\label{eq13}
    A^{(E1M1)} = \frac{8}{27\pi}\alpha^6\int_0^{\Delta}\omega^3(\Delta-\omega)^3\left|S(\omega,\Delta-\omega)\right|^2d\omega
\end{equation}
with $\Delta$ the energy difference between the initial state $E_i$: $(5s^5p)$~$^3$P$_0$ denoted $\ket{i}$ and the final state $E_f$: $(5s^2)$~$^1$S$_0$ denoted $\ket{f}$. This energy difference is equal to 30114\,cm$^{-1}$ and the two photons E1M1 line strength $S(\omega,\Delta-\omega)$ is defined in general by
\begin{equation}
\label{eq14}
\begin{split}
    S(\omega_1,\omega_2) = \sum_{n+}\frac{\bra{f}\boldsymbol{O}^{(M1)}\ket{n+}\bra{n+}\boldsymbol{O}^{(E1)}\ket{i}}{E_{n+}-E_i+\omega_1}\\ +\sum_{n-}\frac{\bra{f}\boldsymbol{O}^{(E1)}\ket{n-}\bra{n-}\boldsymbol{O}^{(M1)}\ket{i}}{E_{n-}-E_i+\omega_2} 
\end{split}
\end{equation}

For bosonic cadmium, the lifetime of the upper $(5s5p)$~$^3$P$_0$ state is consequently estimated to be on the order of 40 years. Table \ref{tab:lifetime} presents a comparison of neutral cadmium decay rates with those of other alkaline-earth-like atoms.

\begin{table}[htbp!]
\centering
\caption{\justifying Comparison of decay rates (in s$^{-1}$) for the cadmium clock transition $(5s^2)$~$^1$S$_0$ $\leftrightarrow$ $(5s5p)$~$^3$P$_0$ clock transition in bosonic and fermionic cadmium with those of other alkaline-earth-like species.}
\label{tab:lifetime}
\begin{ruledtabular}
\begin{tabular}{c|c|c}
                   & Bosons & Fermions \\ \hline

Cd  & 
\makecell{8.8$\times$10$^{-10}$ \\ (present work)} & 
\makecell{4.4$\times$10$^{-2}$ ($^{111}$Cd)~\cite{Garstang1962} \\ 4.8$\times$10$^{-2}$ ($^{113}$Cd)~\cite{Garstang1962}} \\ \hline

Sr~\cite{Santra2004,Porsev2004}  & 
5.5$\times$10$^{-12}$ & 
9$\times$10$^{-3}$  \\ \hline

Ca~\cite{Santra2004,Porsev2004}  & 
3.9$\times$10$^{-13}$ & 
3$\times$10$^{-3}$ \\ \hline

Mg~\cite{Santra2004,Porsev2004}  & 
1.6$\times$10$^{-13}$ & 
9$\times$10$^{-4}$

\end{tabular}
\end{ruledtabular}
\end{table}

\section{C6 long-range dispersion coefficient}
In the study of ultra-cold gases, atomic interactions play a central role and give rise to many of the most striking and useful phenomena~\cite{Weiner1999,Dalfovo1999}. These interactions become important when atoms are cooled below the $\mu$K level, where even weak forces can strongly influence the behavior of the system. To describe these interactions accurately, it is essential to understand the interatomic potential, especially its behavior at large separations where long-range forces dominate. However, while the scattering length—a key quantity characterizing low-energy collisions—is highly sensitive to these long-range parameters, it cannot be reliably determined from them alone~\cite{Gribakin1993}. A meaningful prediction of the scattering length requires not only knowledge of the asymptotic form of the potential but also information about the short-range interactions, where atoms approach closely and exchange becomes significant. Thus, both regions of the potential must be considered together to provide a realistic description of cold collisions.

When considering two identical cadmium atoms in the electronic ground state $\ket{f}$ separated by a distance $R$, their long-range interaction is governed by the leading term of the van der Waals potential, $-C_6/R^6$. The dispersion coefficient $C_6$ plays a central role in the description of low-energy atomic collisions and can be expressed in terms of E1 transition matrix elements~\cite{Margenau1939,OCarroll1968,CHANG1967}:
\begin{equation}
    C_6 = \frac{2}{3}\sum_{i,j}\frac{\big|\bra{i}\boldsymbol{O}^{(E1)}\ket{f}\big|^2\big|\bra{j}\boldsymbol{O}^{(E1)}\ket{f}\big|^2}{E_i+E_j-2E_f},
\end{equation}
where $E_k$ is the energy of a state $\ket{k}$, and the zero of energy is taken at the ground state.

\begin{table}[htbp!]
\caption{\justifying Comparison of $C_6$ coefficients in a.u. for neutral cadmium in the ground state from different sources.}
\label{tab:c6}
\begin{tabular}{cccccccccc}
\toprule \hline
~\cite{Goebel1995}& ~\cite{Koperski2002}a &~\cite{Koperski2002}b&~\cite{Koperski2002}c&~\cite{Moszynski2003}a  &~\cite{Moszynski2003}b & ~\cite{Pahl2011}& ~\cite{Qiao2012} 
 & ~\cite{Yamaguchi2019} & Present  \\
\midrule
466&372 &509 & 611&417&511&840  & 686    & 401(8)         & 395                       \\
\hline
\bottomrule
\end{tabular}%
\end{table}

Table \ref{tab:c6} compares the different values of $C_6$ analytically~\cite{Koperski2002,Moszynski2003,Pahl2011}, via dispersion of the refractive index~\cite{Goebel1995}, and via measurement of the magic wavelength~\cite{Yamaguchi2019}. The present calculation yields $C_6$ = 395\,a.u., in agreement with the value $401$ a.u. reported by Yamaguchi \textit{et al}. This agreement is expected, as both approaches rely on the same fundamental formalism but differ in the determination of the underlying matrix elements. Including contributions from higher-lying states~\cite{Penyazkov2025} modifies the result by less than 1\,$\%$, since the dominant term in the sum arises from the strong E1 coupling between the ground state and the $(5s5p)$~$^1$P$_1$ state.

\section{Conclusion}
We have presented a comprehensive study of the atomic structure and long-range interactions of neutral cadmium using the CI+MBPT framework. General expressions for multipole matrix elements of arbitrary order were derived and applied to evaluate electric and magnetic transition amplitudes, including E1--E3 and M1--M2 processes.

These matrix elements were further employed to compute the lifetimes of the principal transitions relevant to laser cooling and atomic clock applications, including the bosonic clock transition, which has not previously been reported. In addition, they enabled the determination of the long-range van der Waals coefficient $C_6$.

The computed energy levels and E1 matrix elements of the strongest transitions are in agreement with available experimental data, thereby validating the accuracy of the CI+MBPT approach for neutral cadmium. Remarkably, these results were obtained on a standard desktop workstation with 48 GB of RAM, with a total computation time on the order of ten hours.

The accurate characterization of the electronic level structure, together with the precise evaluation of the $C_6$ dispersion coefficient, provides essential groundwork for future investigations of the \textit{s}-wave scattering length in neutral cadmium.

\section*{Acknowledgments}
We would like to thank Yanmei Yu and Dylan Sabulsky for the critical review of the manuscript. 
This work was supported by the European Research Council (Grant No. 772126, ``TICTOCGRAV''). PR and MC acknowledge support from the Horizon Europe program (Grant ID 101080164, ``UVQuanT''), and SM from the Ministero dell’Università e della Ricerca (PRIN 2022 – 2022Z8LX9L, ``ISOTOP'').

\section*{Data Availability}
The data that support the findings of this article are openly available~\cite{data}.

\bibliography{main}

\begin{thebibliography}{79}%
\makeatletter
\providecommand \@ifxundefined [1]{%
 \@ifx{#1\undefined}
}%
\providecommand \@ifnum [1]{%
 \ifnum #1\expandafter \@firstoftwo
 \else \expandafter \@secondoftwo
 \fi
}%
\providecommand \@ifx [1]{%
 \ifx #1\expandafter \@firstoftwo
 \else \expandafter \@secondoftwo
 \fi
}%
\providecommand \natexlab [1]{#1}%
\providecommand \enquote  [1]{``#1''}%
\providecommand \bibnamefont  [1]{#1}%
\providecommand \bibfnamefont [1]{#1}%
\providecommand \citenamefont [1]{#1}%
\providecommand \href@noop [0]{\@secondoftwo}%
\providecommand \href [0]{\begingroup \@sanitize@url \@href}%
\providecommand \@href[1]{\@@startlink{#1}\@@href}%
\providecommand \@@href[1]{\endgroup#1\@@endlink}%
\providecommand \@sanitize@url [0]{\catcode `\\12\catcode `\$12\catcode `\&12\catcode `\#12\catcode `\^12\catcode `\_12\catcode `\%12\relax}%
\providecommand \@@startlink[1]{}%
\providecommand \@@endlink[0]{}%
\providecommand \url  [0]{\begingroup\@sanitize@url \@url }%
\providecommand \@url [1]{\endgroup\@href {#1}{\urlprefix }}%
\providecommand \urlprefix  [0]{URL }%
\providecommand \Eprint [0]{\href }%
\providecommand \doibase [0]{https://doi.org/}%
\providecommand \selectlanguage [0]{\@gobble}%
\providecommand \bibinfo  [0]{\@secondoftwo}%
\providecommand \bibfield  [0]{\@secondoftwo}%
\providecommand \translation [1]{[#1]}%
\providecommand \BibitemOpen [0]{}%
\providecommand \bibitemStop [0]{}%
\providecommand \bibitemNoStop [0]{.\EOS\space}%
\providecommand \EOS [0]{\spacefactor3000\relax}%
\providecommand \BibitemShut  [1]{\csname bibitem#1\endcsname}%
\let\auto@bib@innerbib\@empty
\bibitem [{\citenamefont {Schreck}\ and\ \citenamefont {van Druten}(2021)}]{Schreck2021}%
  \BibitemOpen
  \bibfield  {author} {\bibinfo {author} {\bibfnamefont {F.}~\bibnamefont {Schreck}}\ and\ \bibinfo {author} {\bibfnamefont {K.}~\bibnamefont {van Druten}},\ }\bibfield  {title} {\bibinfo {title} {Laser cooling for quantum gases},\ }\href {https://doi.org/10.1038/s41567-021-01379-w} {\bibfield  {journal} {\bibinfo  {journal} {Nature Physics}\ }\textbf {\bibinfo {volume} {17}},\ \bibinfo {pages} {1296–1304} (\bibinfo {year} {2021})}\BibitemShut {NoStop}%
\bibitem [{\citenamefont {Poli}\ \emph {et~al.}(2013)\citenamefont {Poli}, \citenamefont {Oates}, \citenamefont {Gill},\ and\ \citenamefont {Tino}}]{Poli2013}%
  \BibitemOpen
  \bibfield  {author} {\bibinfo {author} {\bibfnamefont {N.}~\bibnamefont {Poli}}, \bibinfo {author} {\bibfnamefont {C.~W.}\ \bibnamefont {Oates}}, \bibinfo {author} {\bibfnamefont {P.}~\bibnamefont {Gill}},\ and\ \bibinfo {author} {\bibfnamefont {G.~M.}\ \bibnamefont {Tino}},\ }\bibfield  {title} {\bibinfo {title} {Optical atomic clocks},\ }\href {https://doi.org/10.1393/ncr/i2013-10016-0} {\bibfield  {journal} {\bibinfo  {journal} {Rivista del Nuovo Cimento}\ }\textbf {\bibinfo {volume} {36}},\ \bibinfo {pages} {555} (\bibinfo {year} {2013})}\BibitemShut {NoStop}%
\bibitem [{\citenamefont {Safronova}\ \emph {et~al.}(2018)\citenamefont {Safronova}, \citenamefont {Budker}, \citenamefont {DeMille}, \citenamefont {Kimball}, \citenamefont {Derevianko},\ and\ \citenamefont {Clark}}]{Safronova2018}%
  \BibitemOpen
  \bibfield  {author} {\bibinfo {author} {\bibfnamefont {M.}~\bibnamefont {Safronova}}, \bibinfo {author} {\bibfnamefont {D.}~\bibnamefont {Budker}}, \bibinfo {author} {\bibfnamefont {D.}~\bibnamefont {DeMille}}, \bibinfo {author} {\bibfnamefont {D.~F.~J.}\ \bibnamefont {Kimball}}, \bibinfo {author} {\bibfnamefont {A.}~\bibnamefont {Derevianko}},\ and\ \bibinfo {author} {\bibfnamefont {C.~W.}\ \bibnamefont {Clark}},\ }\bibfield  {title} {\bibinfo {title} {Search for new physics with atoms and molecules},\ }\href {https://doi.org/10.1103/revmodphys.90.025008} {\bibfield  {journal} {\bibinfo  {journal} {Reviews of Modern Physics}\ }\textbf {\bibinfo {volume} {90}},\ \bibinfo {pages} {025008} (\bibinfo {year} {2018})}\BibitemShut {NoStop}%
\bibitem [{\citenamefont {Sabulsky}\ \emph {et~al.}(2019)\citenamefont {Sabulsky}, \citenamefont {Dutta}, \citenamefont {Hinds}, \citenamefont {Elder}, \citenamefont {Burrage},\ and\ \citenamefont {Copeland}}]{Sabulsky2019}%
  \BibitemOpen
  \bibfield  {author} {\bibinfo {author} {\bibfnamefont {D.}~\bibnamefont {Sabulsky}}, \bibinfo {author} {\bibfnamefont {I.}~\bibnamefont {Dutta}}, \bibinfo {author} {\bibfnamefont {E.}~\bibnamefont {Hinds}}, \bibinfo {author} {\bibfnamefont {B.}~\bibnamefont {Elder}}, \bibinfo {author} {\bibfnamefont {C.}~\bibnamefont {Burrage}},\ and\ \bibinfo {author} {\bibfnamefont {E.~J.}\ \bibnamefont {Copeland}},\ }\bibfield  {title} {\bibinfo {title} {Experiment to detect dark energy forces using atom interferometry},\ }\href {https://doi.org/10.1103/physrevlett.123.061102} {\bibfield  {journal} {\bibinfo  {journal} {Physical Review Letters}\ }\textbf {\bibinfo {volume} {123}},\ \bibinfo {pages} {061102} (\bibinfo {year} {2019})}\BibitemShut {NoStop}%
\bibitem [{\citenamefont {Tino}\ and\ \citenamefont {Kasevich}(2014)}]{Tino2014}%
  \BibitemOpen
  \bibfield  {author} {\bibinfo {author} {\bibfnamefont {G.~M.}\ \bibnamefont {Tino}}\ and\ \bibinfo {author} {\bibfnamefont {M.~A.}\ \bibnamefont {Kasevich}},\ }\bibinfo {title} {Atom interferometry},\ in\ \href@noop {} {\emph {\bibinfo {booktitle} {Proc. Int. School of Physics ``Enrico Fermi,'' Course CLXXXVIII, Varenna 2013}}}\ (\bibinfo  {publisher} {Società Italiana di Fisica and IOS Press},\ \bibinfo {address} {Amsterdam},\ \bibinfo {year} {2014})\ \bibinfo {note} {lecture notes from the 2013 Varenna Summer School}\BibitemShut {NoStop}%
\bibitem [{\citenamefont {Brewer}\ \emph {et~al.}(2019)\citenamefont {Brewer}, \citenamefont {Chen}, \citenamefont {Hankin}, \citenamefont {Clements}, \citenamefont {Chou}, \citenamefont {Wineland}, \citenamefont {Hume},\ and\ \citenamefont {Leibrandt}}]{Brewer2019}%
  \BibitemOpen
  \bibfield  {author} {\bibinfo {author} {\bibfnamefont {S.}~\bibnamefont {Brewer}}, \bibinfo {author} {\bibfnamefont {J.-S.}\ \bibnamefont {Chen}}, \bibinfo {author} {\bibfnamefont {A.}~\bibnamefont {Hankin}}, \bibinfo {author} {\bibfnamefont {E.}~\bibnamefont {Clements}}, \bibinfo {author} {\bibfnamefont {C.}~\bibnamefont {Chou}}, \bibinfo {author} {\bibfnamefont {D.}~\bibnamefont {Wineland}}, \bibinfo {author} {\bibfnamefont {D.}~\bibnamefont {Hume}},\ and\ \bibinfo {author} {\bibfnamefont {D.}~\bibnamefont {Leibrandt}},\ }\bibfield  {title} {\bibinfo {title} {$^{27}${A}l$^{+}$ quantum-logic clock with a systematic uncertainty below $10^{-18}$},\ }\href {https://doi.org/10.1103/physrevlett.123.033201} {\bibfield  {journal} {\bibinfo  {journal} {Physical Review Letters}\ }\textbf {\bibinfo {volume} {123}},\ \bibinfo {pages} {033201} (\bibinfo {year} {2019})}\BibitemShut {NoStop}%
\bibitem [{\citenamefont {McGrew}\ \emph {et~al.}(2018)\citenamefont {McGrew}, \citenamefont {Zhang}, \citenamefont {Fasano}, \citenamefont {Sch\"{a}ffer}, \citenamefont {Beloy}, \citenamefont {Nicolodi}, \citenamefont {Brown}, \citenamefont {Hinkley}, \citenamefont {Milani}, \citenamefont {Schioppo}, \citenamefont {Yoon},\ and\ \citenamefont {Ludlow}}]{McGrew2018}%
  \BibitemOpen
  \bibfield  {author} {\bibinfo {author} {\bibfnamefont {W.~F.}\ \bibnamefont {McGrew}}, \bibinfo {author} {\bibfnamefont {X.}~\bibnamefont {Zhang}}, \bibinfo {author} {\bibfnamefont {R.~J.}\ \bibnamefont {Fasano}}, \bibinfo {author} {\bibfnamefont {S.~A.}\ \bibnamefont {Sch\"{a}ffer}}, \bibinfo {author} {\bibfnamefont {K.}~\bibnamefont {Beloy}}, \bibinfo {author} {\bibfnamefont {D.}~\bibnamefont {Nicolodi}}, \bibinfo {author} {\bibfnamefont {R.~C.}\ \bibnamefont {Brown}}, \bibinfo {author} {\bibfnamefont {N.}~\bibnamefont {Hinkley}}, \bibinfo {author} {\bibfnamefont {G.}~\bibnamefont {Milani}}, \bibinfo {author} {\bibfnamefont {M.}~\bibnamefont {Schioppo}}, \bibinfo {author} {\bibfnamefont {T.~H.}\ \bibnamefont {Yoon}},\ and\ \bibinfo {author} {\bibfnamefont {A.~D.}\ \bibnamefont {Ludlow}},\ }\bibfield  {title} {\bibinfo {title} {Atomic clock performance enabling geodesy below the centimetre level},\ }\href {https://doi.org/10.1038/s41586-018-0738-2} {\bibfield  {journal} {\bibinfo  {journal} {Nature}\
  }\textbf {\bibinfo {volume} {564}},\ \bibinfo {pages} {87–90} (\bibinfo {year} {2018})}\BibitemShut {NoStop}%
\bibitem [{\citenamefont {Huntemann}\ \emph {et~al.}(2016)\citenamefont {Huntemann}, \citenamefont {Sanner}, \citenamefont {Lipphardt}, \citenamefont {Tamm},\ and\ \citenamefont {Peik}}]{Huntemann2016}%
  \BibitemOpen
  \bibfield  {author} {\bibinfo {author} {\bibfnamefont {N.}~\bibnamefont {Huntemann}}, \bibinfo {author} {\bibfnamefont {C.}~\bibnamefont {Sanner}}, \bibinfo {author} {\bibfnamefont {B.}~\bibnamefont {Lipphardt}}, \bibinfo {author} {\bibfnamefont {C.}~\bibnamefont {Tamm}},\ and\ \bibinfo {author} {\bibfnamefont {E.}~\bibnamefont {Peik}},\ }\bibfield  {title} {\bibinfo {title} {Single-ion atomic clock with 3 x $10^{-18}$ systematic uncertainty},\ }\href {https://doi.org/10.1103/physrevlett.116.063001} {\bibfield  {journal} {\bibinfo  {journal} {Physical Review Letters}\ }\textbf {\bibinfo {volume} {116}},\ \bibinfo {pages} {063001} (\bibinfo {year} {2016})}\BibitemShut {NoStop}%
\bibitem [{\citenamefont {Ushijima}\ \emph {et~al.}(2015)\citenamefont {Ushijima}, \citenamefont {Takamoto}, \citenamefont {Das}, \citenamefont {Ohkubo},\ and\ \citenamefont {Katori}}]{Ushijima2015}%
  \BibitemOpen
  \bibfield  {author} {\bibinfo {author} {\bibfnamefont {I.}~\bibnamefont {Ushijima}}, \bibinfo {author} {\bibfnamefont {M.}~\bibnamefont {Takamoto}}, \bibinfo {author} {\bibfnamefont {M.}~\bibnamefont {Das}}, \bibinfo {author} {\bibfnamefont {T.}~\bibnamefont {Ohkubo}},\ and\ \bibinfo {author} {\bibfnamefont {H.}~\bibnamefont {Katori}},\ }\bibfield  {title} {\bibinfo {title} {Cryogenic optical lattice clocks},\ }\href {https://doi.org/10.1038/nphoton.2015.5} {\bibfield  {journal} {\bibinfo  {journal} {Nature Photonics}\ }\textbf {\bibinfo {volume} {9}},\ \bibinfo {pages} {185–189} (\bibinfo {year} {2015})}\BibitemShut {NoStop}%
\bibitem [{\citenamefont {Bloom}\ \emph {et~al.}(2014)\citenamefont {Bloom}, \citenamefont {Nicholson}, \citenamefont {Williams}, \citenamefont {Campbell}, \citenamefont {Bishof}, \citenamefont {Zhang}, \citenamefont {Zhang}, \citenamefont {Bromley},\ and\ \citenamefont {Ye}}]{Bloom2014}%
  \BibitemOpen
  \bibfield  {author} {\bibinfo {author} {\bibfnamefont {B.~J.}\ \bibnamefont {Bloom}}, \bibinfo {author} {\bibfnamefont {T.~L.}\ \bibnamefont {Nicholson}}, \bibinfo {author} {\bibfnamefont {J.~R.}\ \bibnamefont {Williams}}, \bibinfo {author} {\bibfnamefont {S.~L.}\ \bibnamefont {Campbell}}, \bibinfo {author} {\bibfnamefont {M.}~\bibnamefont {Bishof}}, \bibinfo {author} {\bibfnamefont {X.}~\bibnamefont {Zhang}}, \bibinfo {author} {\bibfnamefont {W.}~\bibnamefont {Zhang}}, \bibinfo {author} {\bibfnamefont {S.~L.}\ \bibnamefont {Bromley}},\ and\ \bibinfo {author} {\bibfnamefont {J.}~\bibnamefont {Ye}},\ }\bibfield  {title} {\bibinfo {title} {An optical lattice clock with accuracy and stability at the $10^{-18}$ level},\ }\href {https://doi.org/10.1038/nature12941} {\bibfield  {journal} {\bibinfo  {journal} {Nature}\ }\textbf {\bibinfo {volume} {506}},\ \bibinfo {pages} {71–75} (\bibinfo {year} {2014})}\BibitemShut {NoStop}%
\bibitem [{\citenamefont {Marshall}\ \emph {et~al.}(2025)\citenamefont {Marshall}, \citenamefont {Castillo}, \citenamefont {Arthur-Dworschack}, \citenamefont {Aeppli}, \citenamefont {Kim}, \citenamefont {Lee}, \citenamefont {Warfield}, \citenamefont {Hinrichs}, \citenamefont {Nardelli}, \citenamefont {Fortier}, \citenamefont {Ye}, \citenamefont {Leibrandt},\ and\ \citenamefont {Hume}}]{Marshall2025}%
  \BibitemOpen
  \bibfield  {author} {\bibinfo {author} {\bibfnamefont {M.~C.}\ \bibnamefont {Marshall}}, \bibinfo {author} {\bibfnamefont {D.~A.~R.}\ \bibnamefont {Castillo}}, \bibinfo {author} {\bibfnamefont {W.~J.}\ \bibnamefont {Arthur-Dworschack}}, \bibinfo {author} {\bibfnamefont {A.}~\bibnamefont {Aeppli}}, \bibinfo {author} {\bibfnamefont {K.}~\bibnamefont {Kim}}, \bibinfo {author} {\bibfnamefont {D.}~\bibnamefont {Lee}}, \bibinfo {author} {\bibfnamefont {W.}~\bibnamefont {Warfield}}, \bibinfo {author} {\bibfnamefont {J.}~\bibnamefont {Hinrichs}}, \bibinfo {author} {\bibfnamefont {N.~V.}\ \bibnamefont {Nardelli}}, \bibinfo {author} {\bibfnamefont {T.~M.}\ \bibnamefont {Fortier}}, \bibinfo {author} {\bibfnamefont {J.}~\bibnamefont {Ye}}, \bibinfo {author} {\bibfnamefont {D.~R.}\ \bibnamefont {Leibrandt}},\ and\ \bibinfo {author} {\bibfnamefont {D.~B.}\ \bibnamefont {Hume}},\ }\bibfield  {title} {\bibinfo {title} {High-stability single-ion clock with 5.5 x $10^{-19}$ systematic uncertainty},\ }\href
  {https://doi.org/10.1103/hb3c-dk28} {\bibfield  {journal} {\bibinfo  {journal} {Physical Review Letters}\ }\textbf {\bibinfo {volume} {135}},\ \bibinfo {pages} {033201} (\bibinfo {year} {2025})}\BibitemShut {NoStop}%
\bibitem [{\citenamefont {Le~Targat}\ \emph {et~al.}(2013)\citenamefont {Le~Targat}, \citenamefont {Lorini}, \citenamefont {Le~Coq}, \citenamefont {Zawada}, \citenamefont {Guéna}, \citenamefont {Abgrall}, \citenamefont {Gurov}, \citenamefont {Rosenbusch}, \citenamefont {Rovera}, \citenamefont {Nagórny}, \citenamefont {Gartman}, \citenamefont {Westergaard}, \citenamefont {Tobar}, \citenamefont {Lours}, \citenamefont {Santarelli}, \citenamefont {Clairon}, \citenamefont {Bize}, \citenamefont {Laurent}, \citenamefont {Lemonde},\ and\ \citenamefont {Lodewyck}}]{LeTargat2013}%
  \BibitemOpen
  \bibfield  {author} {\bibinfo {author} {\bibfnamefont {R.}~\bibnamefont {Le~Targat}}, \bibinfo {author} {\bibfnamefont {L.}~\bibnamefont {Lorini}}, \bibinfo {author} {\bibfnamefont {Y.}~\bibnamefont {Le~Coq}}, \bibinfo {author} {\bibfnamefont {M.}~\bibnamefont {Zawada}}, \bibinfo {author} {\bibfnamefont {J.}~\bibnamefont {Guéna}}, \bibinfo {author} {\bibfnamefont {M.}~\bibnamefont {Abgrall}}, \bibinfo {author} {\bibfnamefont {M.}~\bibnamefont {Gurov}}, \bibinfo {author} {\bibfnamefont {P.}~\bibnamefont {Rosenbusch}}, \bibinfo {author} {\bibfnamefont {D.~G.}\ \bibnamefont {Rovera}}, \bibinfo {author} {\bibfnamefont {B.}~\bibnamefont {Nagórny}}, \bibinfo {author} {\bibfnamefont {R.}~\bibnamefont {Gartman}}, \bibinfo {author} {\bibfnamefont {P.~G.}\ \bibnamefont {Westergaard}}, \bibinfo {author} {\bibfnamefont {M.~E.}\ \bibnamefont {Tobar}}, \bibinfo {author} {\bibfnamefont {M.}~\bibnamefont {Lours}}, \bibinfo {author} {\bibfnamefont {G.}~\bibnamefont {Santarelli}}, \bibinfo {author} {\bibfnamefont
  {A.}~\bibnamefont {Clairon}}, \bibinfo {author} {\bibfnamefont {S.}~\bibnamefont {Bize}}, \bibinfo {author} {\bibfnamefont {P.}~\bibnamefont {Laurent}}, \bibinfo {author} {\bibfnamefont {P.}~\bibnamefont {Lemonde}},\ and\ \bibinfo {author} {\bibfnamefont {J.}~\bibnamefont {Lodewyck}},\ }\bibfield  {title} {\bibinfo {title} {Experimental realization of an optical second with strontium lattice clocks},\ }\href {https://doi.org/10.1038/ncomms3109} {\bibfield  {journal} {\bibinfo  {journal} {Nature Communications}\ }\textbf {\bibinfo {volume} {4}},\ \bibinfo {pages} {2109} (\bibinfo {year} {2013})}\BibitemShut {NoStop}%
\bibitem [{\citenamefont {Huntemann}\ \emph {et~al.}(2014)\citenamefont {Huntemann}, \citenamefont {Lipphardt}, \citenamefont {Tamm}, \citenamefont {Gerginov}, \citenamefont {Weyers},\ and\ \citenamefont {Peik}}]{Huntemann2014}%
  \BibitemOpen
  \bibfield  {author} {\bibinfo {author} {\bibfnamefont {N.}~\bibnamefont {Huntemann}}, \bibinfo {author} {\bibfnamefont {B.}~\bibnamefont {Lipphardt}}, \bibinfo {author} {\bibfnamefont {C.}~\bibnamefont {Tamm}}, \bibinfo {author} {\bibfnamefont {V.}~\bibnamefont {Gerginov}}, \bibinfo {author} {\bibfnamefont {S.}~\bibnamefont {Weyers}},\ and\ \bibinfo {author} {\bibfnamefont {E.}~\bibnamefont {Peik}},\ }\bibfield  {title} {\bibinfo {title} {Improved limit on a temporal variation of m$_p$/m$_e$ from comparisons of {Y}b$^+$ and {C}s atomic clocks},\ }\href {https://doi.org/10.1103/physrevlett.113.210802} {\bibfield  {journal} {\bibinfo  {journal} {Physical Review Letters}\ }\textbf {\bibinfo {volume} {113}},\ \bibinfo {pages} {210802} (\bibinfo {year} {2014})}\BibitemShut {NoStop}%
\bibitem [{\citenamefont {Morel}\ \emph {et~al.}(2020)\citenamefont {Morel}, \citenamefont {Yao}, \citenamefont {Cladé},\ and\ \citenamefont {Guellati-Khélifa}}]{Morel2020}%
  \BibitemOpen
  \bibfield  {author} {\bibinfo {author} {\bibfnamefont {L.}~\bibnamefont {Morel}}, \bibinfo {author} {\bibfnamefont {Z.}~\bibnamefont {Yao}}, \bibinfo {author} {\bibfnamefont {P.}~\bibnamefont {Cladé}},\ and\ \bibinfo {author} {\bibfnamefont {S.}~\bibnamefont {Guellati-Khélifa}},\ }\bibfield  {title} {\bibinfo {title} {Determination of the fine-structure constant with an accuracy of 81 parts per trillion},\ }\href {https://doi.org/10.1038/s41586-020-2964-7} {\bibfield  {journal} {\bibinfo  {journal} {Nature}\ }\textbf {\bibinfo {volume} {588}},\ \bibinfo {pages} {61–65} (\bibinfo {year} {2020})}\BibitemShut {NoStop}%
\bibitem [{\citenamefont {McFerran}\ \emph {et~al.}(2012)\citenamefont {McFerran}, \citenamefont {Yi}, \citenamefont {Mejri}, \citenamefont {Di~Manno}, \citenamefont {Zhang}, \citenamefont {Guéna}, \citenamefont {Le~Coq},\ and\ \citenamefont {Bize}}]{McFerran2012}%
  \BibitemOpen
  \bibfield  {author} {\bibinfo {author} {\bibfnamefont {J.~J.}\ \bibnamefont {McFerran}}, \bibinfo {author} {\bibfnamefont {L.}~\bibnamefont {Yi}}, \bibinfo {author} {\bibfnamefont {S.}~\bibnamefont {Mejri}}, \bibinfo {author} {\bibfnamefont {S.}~\bibnamefont {Di~Manno}}, \bibinfo {author} {\bibfnamefont {W.}~\bibnamefont {Zhang}}, \bibinfo {author} {\bibfnamefont {J.}~\bibnamefont {Guéna}}, \bibinfo {author} {\bibfnamefont {Y.}~\bibnamefont {Le~Coq}},\ and\ \bibinfo {author} {\bibfnamefont {S.}~\bibnamefont {Bize}},\ }\bibfield  {title} {\bibinfo {title} {Neutral atom frequency reference in the deep ultraviolet with fractional uncertainty = 5 x $10^{-15}$},\ }\href {https://doi.org/10.1103/physrevlett.108.183004} {\bibfield  {journal} {\bibinfo  {journal} {Physical Review Letters}\ }\textbf {\bibinfo {volume} {108}},\ \bibinfo {pages} {183004} (\bibinfo {year} {2012})}\BibitemShut {NoStop}%
\bibitem [{\citenamefont {Yamanaka}\ \emph {et~al.}(2015)\citenamefont {Yamanaka}, \citenamefont {Ohmae}, \citenamefont {Ushijima}, \citenamefont {Takamoto},\ and\ \citenamefont {Katori}}]{Yamanaka2015}%
  \BibitemOpen
  \bibfield  {author} {\bibinfo {author} {\bibfnamefont {K.}~\bibnamefont {Yamanaka}}, \bibinfo {author} {\bibfnamefont {N.}~\bibnamefont {Ohmae}}, \bibinfo {author} {\bibfnamefont {I.}~\bibnamefont {Ushijima}}, \bibinfo {author} {\bibfnamefont {M.}~\bibnamefont {Takamoto}},\ and\ \bibinfo {author} {\bibfnamefont {H.}~\bibnamefont {Katori}},\ }\bibfield  {title} {\bibinfo {title} {Frequency ratio of $^{199}${H}g and $^{87}${S}r optical lattice clocks beyond the si limit},\ }\href {https://doi.org/10.1103/physrevlett.114.230801} {\bibfield  {journal} {\bibinfo  {journal} {Physical Review Letters}\ }\textbf {\bibinfo {volume} {114}},\ \bibinfo {pages} {230801} (\bibinfo {year} {2015})}\BibitemShut {NoStop}%
\bibitem [{\citenamefont {Dzuba}\ and\ \citenamefont {Derevianko}(2019)}]{Dzuba2019}%
  \BibitemOpen
  \bibfield  {author} {\bibinfo {author} {\bibfnamefont {V.~A.}\ \bibnamefont {Dzuba}}\ and\ \bibinfo {author} {\bibfnamefont {A.}~\bibnamefont {Derevianko}},\ }\bibfield  {title} {\bibinfo {title} {Blackbody radiation shift for the $^1${S}$_0$ - $^3${P}$_0$ optical clock transition in zinc and cadmium atoms},\ }\href {https://doi.org/10.1088/1361-6455/ab4434} {\bibfield  {journal} {\bibinfo  {journal} {Journal of Physics B: Atomic, Molecular and Optical Physics}\ }\textbf {\bibinfo {volume} {52}},\ \bibinfo {pages} {215005} (\bibinfo {year} {2019})}\BibitemShut {NoStop}%
\bibitem [{\citenamefont {Graham}\ \emph {et~al.}(2013)\citenamefont {Graham}, \citenamefont {Hogan}, \citenamefont {Kasevich},\ and\ \citenamefont {Rajendran}}]{Graham2013}%
  \BibitemOpen
  \bibfield  {author} {\bibinfo {author} {\bibfnamefont {P.~W.}\ \bibnamefont {Graham}}, \bibinfo {author} {\bibfnamefont {J.~M.}\ \bibnamefont {Hogan}}, \bibinfo {author} {\bibfnamefont {M.~A.}\ \bibnamefont {Kasevich}},\ and\ \bibinfo {author} {\bibfnamefont {S.}~\bibnamefont {Rajendran}},\ }\bibfield  {title} {\bibinfo {title} {New method for gravitational wave detection with atomic sensors},\ }\href {https://doi.org/10.1103/physrevlett.110.171102} {\bibfield  {journal} {\bibinfo  {journal} {Physical Review Letters}\ }\textbf {\bibinfo {volume} {110}},\ \bibinfo {pages} {171102} (\bibinfo {year} {2013})}\BibitemShut {NoStop}%
\bibitem [{\citenamefont {Hamilton}\ \emph {et~al.}(2015)\citenamefont {Hamilton}, \citenamefont {Jaffe}, \citenamefont {Haslinger}, \citenamefont {Simmons}, \citenamefont {M\"{u}ller},\ and\ \citenamefont {Khoury}}]{Hamilton2015}%
  \BibitemOpen
  \bibfield  {author} {\bibinfo {author} {\bibfnamefont {P.}~\bibnamefont {Hamilton}}, \bibinfo {author} {\bibfnamefont {M.}~\bibnamefont {Jaffe}}, \bibinfo {author} {\bibfnamefont {P.}~\bibnamefont {Haslinger}}, \bibinfo {author} {\bibfnamefont {Q.}~\bibnamefont {Simmons}}, \bibinfo {author} {\bibfnamefont {H.}~\bibnamefont {M\"{u}ller}},\ and\ \bibinfo {author} {\bibfnamefont {J.}~\bibnamefont {Khoury}},\ }\bibfield  {title} {\bibinfo {title} {Atom-interferometry constraints on dark energy},\ }\href {https://doi.org/10.1126/science.aaa8883} {\bibfield  {journal} {\bibinfo  {journal} {Science}\ }\textbf {\bibinfo {volume} {349}},\ \bibinfo {pages} {849–851} (\bibinfo {year} {2015})}\BibitemShut {NoStop}%
\bibitem [{\citenamefont {Graham}\ \emph {et~al.}(2016)\citenamefont {Graham}, \citenamefont {Kaplan}, \citenamefont {Mardon}, \citenamefont {Rajendran},\ and\ \citenamefont {Terrano}}]{Graham2016}%
  \BibitemOpen
  \bibfield  {author} {\bibinfo {author} {\bibfnamefont {P.~W.}\ \bibnamefont {Graham}}, \bibinfo {author} {\bibfnamefont {D.~E.}\ \bibnamefont {Kaplan}}, \bibinfo {author} {\bibfnamefont {J.}~\bibnamefont {Mardon}}, \bibinfo {author} {\bibfnamefont {S.}~\bibnamefont {Rajendran}},\ and\ \bibinfo {author} {\bibfnamefont {W.~A.}\ \bibnamefont {Terrano}},\ }\bibfield  {title} {\bibinfo {title} {Dark matter direct detection with accelerometers},\ }\href {https://doi.org/10.1103/physrevd.93.075029} {\bibfield  {journal} {\bibinfo  {journal} {Physical Review D}\ }\textbf {\bibinfo {volume} {93}},\ \bibinfo {pages} {075029} (\bibinfo {year} {2016})}\BibitemShut {NoStop}%
\bibitem [{\citenamefont {Matei}\ \emph {et~al.}(2017)\citenamefont {Matei}, \citenamefont {Legero}, \citenamefont {H\"{a}fner}, \citenamefont {Grebing}, \citenamefont {Weyrich}, \citenamefont {Zhang}, \citenamefont {Sonderhouse}, \citenamefont {Robinson}, \citenamefont {Ye}, \citenamefont {Riehle},\ and\ \citenamefont {Sterr}}]{Matei2017}%
  \BibitemOpen
  \bibfield  {author} {\bibinfo {author} {\bibfnamefont {D.}~\bibnamefont {Matei}}, \bibinfo {author} {\bibfnamefont {T.}~\bibnamefont {Legero}}, \bibinfo {author} {\bibfnamefont {S.}~\bibnamefont {H\"{a}fner}}, \bibinfo {author} {\bibfnamefont {C.}~\bibnamefont {Grebing}}, \bibinfo {author} {\bibfnamefont {R.}~\bibnamefont {Weyrich}}, \bibinfo {author} {\bibfnamefont {W.}~\bibnamefont {Zhang}}, \bibinfo {author} {\bibfnamefont {L.}~\bibnamefont {Sonderhouse}}, \bibinfo {author} {\bibfnamefont {J.}~\bibnamefont {Robinson}}, \bibinfo {author} {\bibfnamefont {J.}~\bibnamefont {Ye}}, \bibinfo {author} {\bibfnamefont {F.}~\bibnamefont {Riehle}},\ and\ \bibinfo {author} {\bibfnamefont {U.}~\bibnamefont {Sterr}},\ }\bibfield  {title} {\bibinfo {title} {1.5\,µm lasers with sub-10\,m{H}z linewidth},\ }\href {https://doi.org/10.1103/physrevlett.118.263202} {\bibfield  {journal} {\bibinfo  {journal} {Physical Review Letters}\ }\textbf {\bibinfo {volume} {118}},\ \bibinfo {pages} {263202} (\bibinfo {year}
  {2017})}\BibitemShut {NoStop}%
\bibitem [{\citenamefont {Santra}\ \emph {et~al.}(2005)\citenamefont {Santra}, \citenamefont {Arimondo}, \citenamefont {Ido}, \citenamefont {Greene},\ and\ \citenamefont {Ye}}]{Santra2005}%
  \BibitemOpen
  \bibfield  {author} {\bibinfo {author} {\bibfnamefont {R.}~\bibnamefont {Santra}}, \bibinfo {author} {\bibfnamefont {E.}~\bibnamefont {Arimondo}}, \bibinfo {author} {\bibfnamefont {T.}~\bibnamefont {Ido}}, \bibinfo {author} {\bibfnamefont {C.~H.}\ \bibnamefont {Greene}},\ and\ \bibinfo {author} {\bibfnamefont {J.}~\bibnamefont {Ye}},\ }\bibfield  {title} {\bibinfo {title} {High-accuracy optical clock via three-level coherence in neutral bosonic $^{88}${S}r},\ }\href {https://doi.org/10.1103/physrevlett.94.173002} {\bibfield  {journal} {\bibinfo  {journal} {Physical Review Letters}\ }\textbf {\bibinfo {volume} {94}},\ \bibinfo {pages} {173002} (\bibinfo {year} {2005})}\BibitemShut {NoStop}%
\bibitem [{\citenamefont {Robert}\ and\ \citenamefont {Sabulsky}(2024)}]{Robert2024}%
  \BibitemOpen
  \bibfield  {author} {\bibinfo {author} {\bibfnamefont {P.}~\bibnamefont {Robert}}\ and\ \bibinfo {author} {\bibfnamefont {D.~O.}\ \bibnamefont {Sabulsky}},\ }\bibfield  {title} {\bibinfo {title} {Optically dressed three-level coherence in neutral bosonic alkaline-earth-like species},\ }\href {https://doi.org/10.1103/physrevresearch.6.043268} {\bibfield  {journal} {\bibinfo  {journal} {Physical Review Research}\ }\textbf {\bibinfo {volume} {6}},\ \bibinfo {pages} {043268} (\bibinfo {year} {2024})}\BibitemShut {NoStop}%
\bibitem [{\citenamefont {He}\ \emph {et~al.}(2025)\citenamefont {He}, \citenamefont {Pasquiou}, \citenamefont {Escudero}, \citenamefont {Zhou}, \citenamefont {Borkowski},\ and\ \citenamefont {Schreck}}]{He2025}%
  \BibitemOpen
  \bibfield  {author} {\bibinfo {author} {\bibfnamefont {J.}~\bibnamefont {He}}, \bibinfo {author} {\bibfnamefont {B.}~\bibnamefont {Pasquiou}}, \bibinfo {author} {\bibfnamefont {R.~G.}\ \bibnamefont {Escudero}}, \bibinfo {author} {\bibfnamefont {S.}~\bibnamefont {Zhou}}, \bibinfo {author} {\bibfnamefont {M.}~\bibnamefont {Borkowski}},\ and\ \bibinfo {author} {\bibfnamefont {F.}~\bibnamefont {Schreck}},\ }\bibfield  {title} {\bibinfo {title} {Coherent three-photon excitation of the strontium clock transition},\ }\href {https://doi.org/10.1103/physrevresearch.7.l012050} {\bibfield  {journal} {\bibinfo  {journal} {Physical Review Research}\ }\textbf {\bibinfo {volume} {7}},\ \bibinfo {pages} {l012050} (\bibinfo {year} {2025})}\BibitemShut {NoStop}%
\bibitem [{\citenamefont {Carman}\ \emph {et~al.}(2025)\citenamefont {Carman}, \citenamefont {Rudolph}, \citenamefont {Garber}, \citenamefont {Van~de Graaff}, \citenamefont {Swan}, \citenamefont {Jiang}, \citenamefont {Nantel}, \citenamefont {Abe}, \citenamefont {Barcklay},\ and\ \citenamefont {Hogan}}]{Carman2025}%
  \BibitemOpen
  \bibfield  {author} {\bibinfo {author} {\bibfnamefont {S.~P.}\ \bibnamefont {Carman}}, \bibinfo {author} {\bibfnamefont {J.}~\bibnamefont {Rudolph}}, \bibinfo {author} {\bibfnamefont {B.~E.}\ \bibnamefont {Garber}}, \bibinfo {author} {\bibfnamefont {M.~J.}\ \bibnamefont {Van~de Graaff}}, \bibinfo {author} {\bibfnamefont {H.}~\bibnamefont {Swan}}, \bibinfo {author} {\bibfnamefont {Y.}~\bibnamefont {Jiang}}, \bibinfo {author} {\bibfnamefont {M.}~\bibnamefont {Nantel}}, \bibinfo {author} {\bibfnamefont {M.}~\bibnamefont {Abe}}, \bibinfo {author} {\bibfnamefont {R.~L.}\ \bibnamefont {Barcklay}},\ and\ \bibinfo {author} {\bibfnamefont {J.~M.}\ \bibnamefont {Hogan}},\ }\bibfield  {title} {\bibinfo {title} {Collinear three-photon excitation of a strongly forbidden optical clock transition},\ }\href {https://doi.org/10.1103/qk3v-46y8} {\bibfield  {journal} {\bibinfo  {journal} {Physical Review X}\ }\textbf {\bibinfo {volume} {15}},\ \bibinfo {pages} {031051} (\bibinfo {year} {2025})}\BibitemShut {NoStop}%
\bibitem [{\citenamefont {Penyazkov}\ \emph {et~al.}(2025)\citenamefont {Penyazkov}, \citenamefont {Yu}, \citenamefont {Skripnikov},\ and\ \citenamefont {Ding}}]{Penyazkov2025}%
  \BibitemOpen
  \bibfield  {author} {\bibinfo {author} {\bibfnamefont {G.}~\bibnamefont {Penyazkov}}, \bibinfo {author} {\bibfnamefont {Y.}~\bibnamefont {Yu}}, \bibinfo {author} {\bibfnamefont {L.~V.}\ \bibnamefont {Skripnikov}},\ and\ \bibinfo {author} {\bibfnamefont {S.}~\bibnamefont {Ding}},\ }\bibfield  {title} {\bibinfo {title} {Theoretical study of transition matrix elements in cadmium for vacuum-ultraviolet generation in $^{229}${T}h nuclear clock applications},\ }\href {https://doi.org/10.1103/q6b3-mcfp} {\bibfield  {journal} {\bibinfo  {journal} {Physical Review A}\ }\textbf {\bibinfo {volume} {112}},\ \bibinfo {pages} {022807} (\bibinfo {year} {2025})}\BibitemShut {NoStop}%
\bibitem [{\citenamefont {Zhang}\ \emph {et~al.}(2024{\natexlab{a}})\citenamefont {Zhang}, \citenamefont {Ooi}, \citenamefont {Higgins}, \citenamefont {Doyle}, \citenamefont {von~der Wense}, \citenamefont {Beeks}, \citenamefont {Leitner}, \citenamefont {Kazakov}, \citenamefont {Li}, \citenamefont {Thirolf}, \citenamefont {Schumm},\ and\ \citenamefont {Ye}}]{Zhang2024b}%
  \BibitemOpen
  \bibfield  {author} {\bibinfo {author} {\bibfnamefont {C.}~\bibnamefont {Zhang}}, \bibinfo {author} {\bibfnamefont {T.}~\bibnamefont {Ooi}}, \bibinfo {author} {\bibfnamefont {J.~S.}\ \bibnamefont {Higgins}}, \bibinfo {author} {\bibfnamefont {J.~F.}\ \bibnamefont {Doyle}}, \bibinfo {author} {\bibfnamefont {L.}~\bibnamefont {von~der Wense}}, \bibinfo {author} {\bibfnamefont {K.}~\bibnamefont {Beeks}}, \bibinfo {author} {\bibfnamefont {A.}~\bibnamefont {Leitner}}, \bibinfo {author} {\bibfnamefont {G.~A.}\ \bibnamefont {Kazakov}}, \bibinfo {author} {\bibfnamefont {P.}~\bibnamefont {Li}}, \bibinfo {author} {\bibfnamefont {P.~G.}\ \bibnamefont {Thirolf}}, \bibinfo {author} {\bibfnamefont {T.}~\bibnamefont {Schumm}},\ and\ \bibinfo {author} {\bibfnamefont {J.}~\bibnamefont {Ye}},\ }\bibfield  {title} {\bibinfo {title} {Frequency ratio of the 229m{T}h nuclear isomeric transition and the 87{S}r atomic clock},\ }\href {https://doi.org/10.1038/s41586-024-07839-6} {\bibfield  {journal} {\bibinfo  {journal} {Nature}\
  }\textbf {\bibinfo {volume} {633}},\ \bibinfo {pages} {63–70} (\bibinfo {year} {2024}{\natexlab{a}})}\BibitemShut {NoStop}%
\bibitem [{\citenamefont {Hur}\ \emph {et~al.}(2022)\citenamefont {Hur}, \citenamefont {Aude~Craik}, \citenamefont {Counts}, \citenamefont {Knyazev}, \citenamefont {Caldwell}, \citenamefont {Leung}, \citenamefont {Pandey}, \citenamefont {Berengut}, \citenamefont {Geddes}, \citenamefont {Nazarewicz}, \citenamefont {Reinhard}, \citenamefont {Kawasaki}, \citenamefont {Jeon}, \citenamefont {Jhe},\ and\ \citenamefont {Vuletić}}]{Hur2022}%
  \BibitemOpen
  \bibfield  {author} {\bibinfo {author} {\bibfnamefont {J.}~\bibnamefont {Hur}}, \bibinfo {author} {\bibfnamefont {D.~P.}\ \bibnamefont {Aude~Craik}}, \bibinfo {author} {\bibfnamefont {I.}~\bibnamefont {Counts}}, \bibinfo {author} {\bibfnamefont {E.}~\bibnamefont {Knyazev}}, \bibinfo {author} {\bibfnamefont {L.}~\bibnamefont {Caldwell}}, \bibinfo {author} {\bibfnamefont {C.}~\bibnamefont {Leung}}, \bibinfo {author} {\bibfnamefont {S.}~\bibnamefont {Pandey}}, \bibinfo {author} {\bibfnamefont {J.~C.}\ \bibnamefont {Berengut}}, \bibinfo {author} {\bibfnamefont {A.}~\bibnamefont {Geddes}}, \bibinfo {author} {\bibfnamefont {W.}~\bibnamefont {Nazarewicz}}, \bibinfo {author} {\bibfnamefont {P.-G.}\ \bibnamefont {Reinhard}}, \bibinfo {author} {\bibfnamefont {A.}~\bibnamefont {Kawasaki}}, \bibinfo {author} {\bibfnamefont {H.}~\bibnamefont {Jeon}}, \bibinfo {author} {\bibfnamefont {W.}~\bibnamefont {Jhe}},\ and\ \bibinfo {author} {\bibfnamefont {V.}~\bibnamefont {Vuletić}},\ }\bibfield  {title} {\bibinfo {title}
  {Evidence of {T}wo-{S}ource {K}ing {P}lot {N}onlinearity in {S}pectroscopic {S}earch for {N}ew {B}oson},\ }\href {https://doi.org/10.1103/physrevlett.128.163201} {\bibfield  {journal} {\bibinfo  {journal} {Physical Review Letters}\ }\textbf {\bibinfo {volume} {128}},\ \bibinfo {pages} {163201} (\bibinfo {year} {2022})}\BibitemShut {NoStop}%
\bibitem [{\citenamefont {Ohayon}\ \emph {et~al.}(2022)\citenamefont {Ohayon}, \citenamefont {Hofs\"{a}ss}, \citenamefont {Padilla-Castillo}, \citenamefont {Wright}, \citenamefont {Meijer}, \citenamefont {Truppe}, \citenamefont {Gibble},\ and\ \citenamefont {Sahoo}}]{Ohayon2022}%
  \BibitemOpen
  \bibfield  {author} {\bibinfo {author} {\bibfnamefont {B.}~\bibnamefont {Ohayon}}, \bibinfo {author} {\bibfnamefont {S.}~\bibnamefont {Hofs\"{a}ss}}, \bibinfo {author} {\bibfnamefont {J.~E.}\ \bibnamefont {Padilla-Castillo}}, \bibinfo {author} {\bibfnamefont {S.~C.}\ \bibnamefont {Wright}}, \bibinfo {author} {\bibfnamefont {G.}~\bibnamefont {Meijer}}, \bibinfo {author} {\bibfnamefont {S.}~\bibnamefont {Truppe}}, \bibinfo {author} {\bibfnamefont {K.}~\bibnamefont {Gibble}},\ and\ \bibinfo {author} {\bibfnamefont {B.~K.}\ \bibnamefont {Sahoo}},\ }\bibfield  {title} {\bibinfo {title} {Isotope shifts in cadmium as a sensitive probe for physics beyond the standard model},\ }\href {https://doi.org/10.1088/1367-2630/acacbb} {\bibfield  {journal} {\bibinfo  {journal} {New Journal of Physics}\ }\textbf {\bibinfo {volume} {24}},\ \bibinfo {pages} {123040} (\bibinfo {year} {2022})}\BibitemShut {NoStop}%
\bibitem [{\citenamefont {Garstang}(1962)}]{Garstang1962}%
  \BibitemOpen
  \bibfield  {author} {\bibinfo {author} {\bibfnamefont {R.~H.}\ \bibnamefont {Garstang}},\ }\bibfield  {title} {\bibinfo {title} {Hyperfine structure and intercombination line intensities in the spectra of magnesium, zinc, cadmium, and mercury},\ }\href {https://doi.org/10.1364/josa.52.000845} {\bibfield  {journal} {\bibinfo  {journal} {Journal of the Optical Society of America}\ }\textbf {\bibinfo {volume} {52}},\ \bibinfo {pages} {000845} (\bibinfo {year} {1962})}\BibitemShut {NoStop}%
\bibitem [{\citenamefont {Brickman}\ \emph {et~al.}(2007)\citenamefont {Brickman}, \citenamefont {Chang}, \citenamefont {Acton}, \citenamefont {Chew}, \citenamefont {Matsukevich}, \citenamefont {Haljan}, \citenamefont {Bagnato},\ and\ \citenamefont {Monroe}}]{Brickman2007}%
  \BibitemOpen
  \bibfield  {author} {\bibinfo {author} {\bibfnamefont {K.-A.}\ \bibnamefont {Brickman}}, \bibinfo {author} {\bibfnamefont {M.-S.}\ \bibnamefont {Chang}}, \bibinfo {author} {\bibfnamefont {M.}~\bibnamefont {Acton}}, \bibinfo {author} {\bibfnamefont {A.}~\bibnamefont {Chew}}, \bibinfo {author} {\bibfnamefont {D.}~\bibnamefont {Matsukevich}}, \bibinfo {author} {\bibfnamefont {P.~C.}\ \bibnamefont {Haljan}}, \bibinfo {author} {\bibfnamefont {V.~S.}\ \bibnamefont {Bagnato}},\ and\ \bibinfo {author} {\bibfnamefont {C.}~\bibnamefont {Monroe}},\ }\bibfield  {title} {\bibinfo {title} {Magneto-optical trapping of cadmium},\ }\href {https://doi.org/10.1103/physreva.76.043411} {\bibfield  {journal} {\bibinfo  {journal} {Physical Review A}\ }\textbf {\bibinfo {volume} {76}},\ \bibinfo {pages} {043411} (\bibinfo {year} {2007})}\BibitemShut {NoStop}%
\bibitem [{\citenamefont {Bandarupally}\ \emph {et~al.}(2023)\citenamefont {Bandarupally}, \citenamefont {Tinsley}, \citenamefont {Chiarotti},\ and\ \citenamefont {Poli}}]{Bandarupally2023}%
  \BibitemOpen
  \bibfield  {author} {\bibinfo {author} {\bibfnamefont {S.}~\bibnamefont {Bandarupally}}, \bibinfo {author} {\bibfnamefont {J.~N.}\ \bibnamefont {Tinsley}}, \bibinfo {author} {\bibfnamefont {M.}~\bibnamefont {Chiarotti}},\ and\ \bibinfo {author} {\bibfnamefont {N.}~\bibnamefont {Poli}},\ }\bibfield  {title} {\bibinfo {title} {Design and simulation of a source of cold cadmium for atom interferometry},\ }\href {https://doi.org/10.1088/1361-6455/acf3bf} {\bibfield  {journal} {\bibinfo  {journal} {Journal of Physics B: Atomic, Molecular and Optical Physics}\ }\textbf {\bibinfo {volume} {56}},\ \bibinfo {pages} {185301} (\bibinfo {year} {2023})}\BibitemShut {NoStop}%
\bibitem [{\citenamefont {Padilla-Castillo}\ \emph {et~al.}(2025)\citenamefont {Padilla-Castillo}, \citenamefont {Hofs\"{a}ss}, \citenamefont {Palánki}, \citenamefont {Cai}, \citenamefont {Rich}, \citenamefont {Thomas}, \citenamefont {Kray}, \citenamefont {Meijer}, \citenamefont {Wright},\ and\ \citenamefont {Truppe}}]{PadillaCastilloarxiv}%
  \BibitemOpen
  \bibfield  {author} {\bibinfo {author} {\bibfnamefont {J.~E.}\ \bibnamefont {Padilla-Castillo}}, \bibinfo {author} {\bibfnamefont {S.}~\bibnamefont {Hofs\"{a}ss}}, \bibinfo {author} {\bibfnamefont {L.}~\bibnamefont {Palánki}}, \bibinfo {author} {\bibfnamefont {J.}~\bibnamefont {Cai}}, \bibinfo {author} {\bibfnamefont {C.~J.~H.}\ \bibnamefont {Rich}}, \bibinfo {author} {\bibfnamefont {R.}~\bibnamefont {Thomas}}, \bibinfo {author} {\bibfnamefont {S.}~\bibnamefont {Kray}}, \bibinfo {author} {\bibfnamefont {G.}~\bibnamefont {Meijer}}, \bibinfo {author} {\bibfnamefont {S.~C.}\ \bibnamefont {Wright}},\ and\ \bibinfo {author} {\bibfnamefont {S.}~\bibnamefont {Truppe}},\ }\bibfield  {title} {\bibinfo {title} {A large magneto-optical trap of cadmium atoms loaded from a cryogenic buffer gas beam},\ }\href {https://arxiv.org/abs/2506.01180} {\bibfield  {journal} {\bibinfo  {journal} {arXiv:2506.01180}\ } (\bibinfo {year} {2025})}\BibitemShut {NoStop}%
\bibitem [{\citenamefont {Walker}\ \emph {et~al.}(1990)\citenamefont {Walker}, \citenamefont {Sesko},\ and\ \citenamefont {Wieman}}]{Walker1990}%
  \BibitemOpen
  \bibfield  {author} {\bibinfo {author} {\bibfnamefont {T.}~\bibnamefont {Walker}}, \bibinfo {author} {\bibfnamefont {D.}~\bibnamefont {Sesko}},\ and\ \bibinfo {author} {\bibfnamefont {C.}~\bibnamefont {Wieman}},\ }\bibfield  {title} {\bibinfo {title} {Collective behavior of optically trapped neutral atoms},\ }\href {https://doi.org/10.1103/physrevlett.64.408} {\bibfield  {journal} {\bibinfo  {journal} {Physical Review Letters}\ }\textbf {\bibinfo {volume} {64}},\ \bibinfo {pages} {408–411} (\bibinfo {year} {1990})}\BibitemShut {NoStop}%
\bibitem [{\citenamefont {Duarte}\ \emph {et~al.}(2011)\citenamefont {Duarte}, \citenamefont {Hart}, \citenamefont {Hitchcock}, \citenamefont {Corcovilos}, \citenamefont {Yang}, \citenamefont {Reed},\ and\ \citenamefont {Hulet}}]{Duarte2011}%
  \BibitemOpen
  \bibfield  {author} {\bibinfo {author} {\bibfnamefont {P.~M.}\ \bibnamefont {Duarte}}, \bibinfo {author} {\bibfnamefont {R.~A.}\ \bibnamefont {Hart}}, \bibinfo {author} {\bibfnamefont {J.~M.}\ \bibnamefont {Hitchcock}}, \bibinfo {author} {\bibfnamefont {T.~A.}\ \bibnamefont {Corcovilos}}, \bibinfo {author} {\bibfnamefont {T.-L.}\ \bibnamefont {Yang}}, \bibinfo {author} {\bibfnamefont {A.}~\bibnamefont {Reed}},\ and\ \bibinfo {author} {\bibfnamefont {R.~G.}\ \bibnamefont {Hulet}},\ }\bibfield  {title} {\bibinfo {title} {All-optical production of a lithium quantum gas using narrow-line laser cooling},\ }\href {https://doi.org/10.1103/physreva.84.061406} {\bibfield  {journal} {\bibinfo  {journal} {Physical Review A}\ }\textbf {\bibinfo {volume} {84}},\ \bibinfo {pages} {061406} (\bibinfo {year} {2011})}\BibitemShut {NoStop}%
\bibitem [{\citenamefont {Stellmer}\ \emph {et~al.}(2013)\citenamefont {Stellmer}, \citenamefont {Pasquiou}, \citenamefont {Grimm},\ and\ \citenamefont {Schreck}}]{Stellmer2013}%
  \BibitemOpen
  \bibfield  {author} {\bibinfo {author} {\bibfnamefont {S.}~\bibnamefont {Stellmer}}, \bibinfo {author} {\bibfnamefont {B.}~\bibnamefont {Pasquiou}}, \bibinfo {author} {\bibfnamefont {R.}~\bibnamefont {Grimm}},\ and\ \bibinfo {author} {\bibfnamefont {F.}~\bibnamefont {Schreck}},\ }\bibfield  {title} {\bibinfo {title} {Laser cooling to quantum degeneracy},\ }\href {https://doi.org/10.1103/physrevlett.110.263003} {\bibfield  {journal} {\bibinfo  {journal} {Physical Review Letters}\ }\textbf {\bibinfo {volume} {110}},\ \bibinfo {pages} {263003} (\bibinfo {year} {2013})}\BibitemShut {NoStop}%
\bibitem [{\citenamefont {Chen}\ \emph {et~al.}(2022)\citenamefont {Chen}, \citenamefont {González~Escudero}, \citenamefont {Minář}, \citenamefont {Pasquiou}, \citenamefont {Bennetts},\ and\ \citenamefont {Schreck}}]{Chen2022}%
  \BibitemOpen
  \bibfield  {author} {\bibinfo {author} {\bibfnamefont {C.-C.}\ \bibnamefont {Chen}}, \bibinfo {author} {\bibfnamefont {R.}~\bibnamefont {González~Escudero}}, \bibinfo {author} {\bibfnamefont {J.}~\bibnamefont {Minář}}, \bibinfo {author} {\bibfnamefont {B.}~\bibnamefont {Pasquiou}}, \bibinfo {author} {\bibfnamefont {S.}~\bibnamefont {Bennetts}},\ and\ \bibinfo {author} {\bibfnamefont {F.}~\bibnamefont {Schreck}},\ }\bibfield  {title} {\bibinfo {title} {Continuous {B}ose–{E}instein condensation},\ }\href {https://doi.org/10.1038/s41586-022-04731-z} {\bibfield  {journal} {\bibinfo  {journal} {Nature}\ }\textbf {\bibinfo {volume} {606}},\ \bibinfo {pages} {683–687} (\bibinfo {year} {2022})}\BibitemShut {NoStop}%
\bibitem [{\citenamefont {Kaneda}\ \emph {et~al.}(2016)\citenamefont {Kaneda}, \citenamefont {Yarborough}, \citenamefont {Merzlyak}, \citenamefont {Yamaguchi}, \citenamefont {Hayashida}, \citenamefont {Ohmae},\ and\ \citenamefont {Katori}}]{Kaneda2016}%
  \BibitemOpen
  \bibfield  {author} {\bibinfo {author} {\bibfnamefont {Y.}~\bibnamefont {Kaneda}}, \bibinfo {author} {\bibfnamefont {J.~M.}\ \bibnamefont {Yarborough}}, \bibinfo {author} {\bibfnamefont {Y.}~\bibnamefont {Merzlyak}}, \bibinfo {author} {\bibfnamefont {A.}~\bibnamefont {Yamaguchi}}, \bibinfo {author} {\bibfnamefont {K.}~\bibnamefont {Hayashida}}, \bibinfo {author} {\bibfnamefont {N.}~\bibnamefont {Ohmae}},\ and\ \bibinfo {author} {\bibfnamefont {H.}~\bibnamefont {Katori}},\ }\bibfield  {title} {\bibinfo {title} {Continuous-wave, single-frequency 229 nm laser source for laser cooling of cadmium atoms},\ }\href {https://doi.org/10.1364/ol.41.000705} {\bibfield  {journal} {\bibinfo  {journal} {Optics Letters}\ }\textbf {\bibinfo {volume} {41}},\ \bibinfo {pages} {705} (\bibinfo {year} {2016})}\BibitemShut {NoStop}%
\bibitem [{\citenamefont {Yamaguchi}\ \emph {et~al.}(2019)\citenamefont {Yamaguchi}, \citenamefont {Safronova}, \citenamefont {Gibble},\ and\ \citenamefont {Katori}}]{Yamaguchi2019}%
  \BibitemOpen
  \bibfield  {author} {\bibinfo {author} {\bibfnamefont {A.}~\bibnamefont {Yamaguchi}}, \bibinfo {author} {\bibfnamefont {M.}~\bibnamefont {Safronova}}, \bibinfo {author} {\bibfnamefont {K.}~\bibnamefont {Gibble}},\ and\ \bibinfo {author} {\bibfnamefont {H.}~\bibnamefont {Katori}},\ }\bibfield  {title} {\bibinfo {title} {Narrow-line cooling and determination of the magic wavelength of {C}d},\ }\href {https://doi.org/10.1103/physrevlett.123.113201} {\bibfield  {journal} {\bibinfo  {journal} {Physical Review Letters}\ }\textbf {\bibinfo {volume} {123}},\ \bibinfo {pages} {113201} (\bibinfo {year} {2019})}\BibitemShut {NoStop}%
\bibitem [{\citenamefont {Tinsley}\ \emph {et~al.}(2022)\citenamefont {Tinsley}, \citenamefont {Bandarupally}, \citenamefont {Chiarotti}, \citenamefont {Manzoor}, \citenamefont {Salvi},\ and\ \citenamefont {Poli}}]{Tinsley2022}%
  \BibitemOpen
  \bibfield  {author} {\bibinfo {author} {\bibfnamefont {J.}~\bibnamefont {Tinsley}}, \bibinfo {author} {\bibfnamefont {S.}~\bibnamefont {Bandarupally}}, \bibinfo {author} {\bibfnamefont {M.}~\bibnamefont {Chiarotti}}, \bibinfo {author} {\bibfnamefont {S.}~\bibnamefont {Manzoor}}, \bibinfo {author} {\bibfnamefont {L.}~\bibnamefont {Salvi}},\ and\ \bibinfo {author} {\bibfnamefont {N.}~\bibnamefont {Poli}},\ }\bibfield  {title} {\bibinfo {title} {Prospects for a simultaneous atom interferometer with ultracold cadmium and strontium for fundamental physics tests},\ }in\ \href {https://doi.org/10.1117/12.2616918} {\emph {\bibinfo {booktitle} {Optical and Quantum Sensing and Precision Metrology II}}},\ \bibinfo {editor} {edited by\ \bibinfo {editor} {\bibfnamefont {S.~M.}\ \bibnamefont {Shahriar}}\ and\ \bibinfo {editor} {\bibfnamefont {J.}~\bibnamefont {Scheuer}}}\ (\bibinfo  {publisher} {SPIE},\ \bibinfo {year} {2022})\ p.~\bibinfo {pages} {4}\BibitemShut {NoStop}%
\bibitem [{\citenamefont {Moszynski}\ \emph {et~al.}(2003)\citenamefont {Moszynski}, \citenamefont {Łach}, \citenamefont {Jaszuński},\ and\ \citenamefont {Bussery-Honvault}}]{Moszynski2003}%
  \BibitemOpen
  \bibfield  {author} {\bibinfo {author} {\bibfnamefont {R.}~\bibnamefont {Moszynski}}, \bibinfo {author} {\bibfnamefont {G.}~\bibnamefont {Łach}}, \bibinfo {author} {\bibfnamefont {M.}~\bibnamefont {Jaszuński}},\ and\ \bibinfo {author} {\bibfnamefont {B.}~\bibnamefont {Bussery-Honvault}},\ }\bibfield  {title} {\bibinfo {title} {Long-range relativistic interactions in the {C}owan-{G}riffin approximation and their {QED} retardation: Application to helium, calcium, and cadmium dimers},\ }\href {https://doi.org/10.1103/physreva.68.052706} {\bibfield  {journal} {\bibinfo  {journal} {Physical Review A}\ }\textbf {\bibinfo {volume} {68}},\ \bibinfo {pages} {052706} (\bibinfo {year} {2003})}\BibitemShut {NoStop}%
\bibitem [{\citenamefont {Porsev}\ and\ \citenamefont {Safronova}(2020)}]{Porsev2020}%
  \BibitemOpen
  \bibfield  {author} {\bibinfo {author} {\bibfnamefont {S.~G.}\ \bibnamefont {Porsev}}\ and\ \bibinfo {author} {\bibfnamefont {M.~S.}\ \bibnamefont {Safronova}},\ }\bibfield  {title} {\bibinfo {title} {Calculation of higher-order corrections to the light shift of the 5s$^2$ $^1${S}$_0$ - 5s5p $^3${P}$_0$ clock transition in {C}d},\ }\href {https://doi.org/10.1103/physreva.102.012811} {\bibfield  {journal} {\bibinfo  {journal} {Physical Review A}\ }\textbf {\bibinfo {volume} {102}},\ \bibinfo {pages} {012811} (\bibinfo {year} {2020})}\BibitemShut {NoStop}%
\bibitem [{\citenamefont {Zhang}\ \emph {et~al.}(2024{\natexlab{b}})\citenamefont {Zhang}, \citenamefont {Jiang}, \citenamefont {Dong},\ and\ \citenamefont {Tang}}]{Zhang2024}%
  \BibitemOpen
  \bibfield  {author} {\bibinfo {author} {\bibfnamefont {R.-K.}\ \bibnamefont {Zhang}}, \bibinfo {author} {\bibfnamefont {J.}~\bibnamefont {Jiang}}, \bibinfo {author} {\bibfnamefont {C.-Z.}\ \bibnamefont {Dong}},\ and\ \bibinfo {author} {\bibfnamefont {Y.-B.}\ \bibnamefont {Tang}},\ }\bibfield  {title} {\bibinfo {title} {Dynamic polarizabilities and triply magic trapping conditions for 5s$^{2}$ $^1{S}_0$ $\rightarrow$ $5s5p$ $^{3}{P}_{0,2}$ transitions of {C}d atoms},\ }\href {https://doi.org/10.1103/physreva.109.032821} {\bibfield  {journal} {\bibinfo  {journal} {Physical Review A}\ }\textbf {\bibinfo {volume} {109}},\ \bibinfo {pages} {032821} (\bibinfo {year} {2024}{\natexlab{b}})}\BibitemShut {NoStop}%
\bibitem [{\citenamefont {Schelfhout}\ and\ \citenamefont {McFerran}(2022)}]{Schelfhout2022}%
  \BibitemOpen
  \bibfield  {author} {\bibinfo {author} {\bibfnamefont {J.~S.}\ \bibnamefont {Schelfhout}}\ and\ \bibinfo {author} {\bibfnamefont {J.~J.}\ \bibnamefont {McFerran}},\ }\bibfield  {title} {\bibinfo {title} {Multiconfiguration {D}irac-{H}artree-{F}ock calculations for {H}g and {C}d with estimates for unknown clock-transition frequencies},\ }\href {https://doi.org/10.1103/physreva.105.022805} {\bibfield  {journal} {\bibinfo  {journal} {Physical Review A}\ }\textbf {\bibinfo {volume} {105}},\ \bibinfo {pages} {022805} (\bibinfo {year} {2022})}\BibitemShut {NoStop}%
\bibitem [{\citenamefont {Berengut}(2016)}]{Berengut2016}%
  \BibitemOpen
  \bibfield  {author} {\bibinfo {author} {\bibfnamefont {J.~C.}\ \bibnamefont {Berengut}},\ }\bibfield  {title} {\bibinfo {title} {Particle-hole configuration interaction and many-body perturbation theory: Application to {H}g$^+$},\ }\href {https://doi.org/10.1103/physreva.94.012502} {\bibfield  {journal} {\bibinfo  {journal} {Physical Review A}\ }\textbf {\bibinfo {volume} {94}},\ \bibinfo {pages} {012502} (\bibinfo {year} {2016})}\BibitemShut {NoStop}%
\bibitem [{\citenamefont {Torretti}\ \emph {et~al.}(2017)\citenamefont {Torretti}, \citenamefont {Windberger}, \citenamefont {Ryabtsev}, \citenamefont {Dobrodey}, \citenamefont {Bekker}, \citenamefont {Ubachs}, \citenamefont {Hoekstra}, \citenamefont {Kahl}, \citenamefont {Berengut}, \citenamefont {López-Urrutia},\ and\ \citenamefont {Versolato}}]{Torretti2017}%
  \BibitemOpen
  \bibfield  {author} {\bibinfo {author} {\bibfnamefont {F.}~\bibnamefont {Torretti}}, \bibinfo {author} {\bibfnamefont {A.}~\bibnamefont {Windberger}}, \bibinfo {author} {\bibfnamefont {A.}~\bibnamefont {Ryabtsev}}, \bibinfo {author} {\bibfnamefont {S.}~\bibnamefont {Dobrodey}}, \bibinfo {author} {\bibfnamefont {H.}~\bibnamefont {Bekker}}, \bibinfo {author} {\bibfnamefont {W.}~\bibnamefont {Ubachs}}, \bibinfo {author} {\bibfnamefont {R.}~\bibnamefont {Hoekstra}}, \bibinfo {author} {\bibfnamefont {E.~V.}\ \bibnamefont {Kahl}}, \bibinfo {author} {\bibfnamefont {J.~C.}\ \bibnamefont {Berengut}}, \bibinfo {author} {\bibfnamefont {J.~R.~C.}\ \bibnamefont {López-Urrutia}},\ and\ \bibinfo {author} {\bibfnamefont {O.~O.}\ \bibnamefont {Versolato}},\ }\bibfield  {title} {\bibinfo {title} {Optical spectroscopy of complex open-4d-shell ions {S}n$^{7+}$–{S}n$^{10+}$},\ }\href {https://doi.org/10.1103/physreva.95.042503} {\bibfield  {journal} {\bibinfo  {journal} {Physical Review A}\ }\textbf {\bibinfo {volume}
  {95}},\ \bibinfo {pages} {042503} (\bibinfo {year} {2017})}\BibitemShut {NoStop}%
\bibitem [{\citenamefont {Berengut}\ \emph {et~al.}(2006)\citenamefont {Berengut}, \citenamefont {Flambaum},\ and\ \citenamefont {Kozlov}}]{Berengut2006}%
  \BibitemOpen
  \bibfield  {author} {\bibinfo {author} {\bibfnamefont {J.~C.}\ \bibnamefont {Berengut}}, \bibinfo {author} {\bibfnamefont {V.~V.}\ \bibnamefont {Flambaum}},\ and\ \bibinfo {author} {\bibfnamefont {M.~G.}\ \bibnamefont {Kozlov}},\ }\bibfield  {title} {\bibinfo {title} {Calculation of isotope shifts and relativistic shifts in {C} {I}, {C} {II}, {C} {III}, and {C} {IV}},\ }\href {https://doi.org/10.1103/physreva.73.012504} {\bibfield  {journal} {\bibinfo  {journal} {Physical Review A}\ }\textbf {\bibinfo {volume} {73}},\ \bibinfo {pages} {012504} (\bibinfo {year} {2006})}\BibitemShut {NoStop}%
\bibitem [{\citenamefont {Dzuba}\ \emph {et~al.}(1996)\citenamefont {Dzuba}, \citenamefont {Flambaum},\ and\ \citenamefont {Kozlov}}]{Dzuba1996}%
  \BibitemOpen
  \bibfield  {author} {\bibinfo {author} {\bibfnamefont {V.~A.}\ \bibnamefont {Dzuba}}, \bibinfo {author} {\bibfnamefont {V.~V.}\ \bibnamefont {Flambaum}},\ and\ \bibinfo {author} {\bibfnamefont {M.~G.}\ \bibnamefont {Kozlov}},\ }\bibfield  {title} {\bibinfo {title} {Combination of the many-body perturbation theory with the configuration-interaction method},\ }\href {https://doi.org/10.1103/physreva.54.3948} {\bibfield  {journal} {\bibinfo  {journal} {Physical Review A}\ }\textbf {\bibinfo {volume} {54}},\ \bibinfo {pages} {3948–3959} (\bibinfo {year} {1996})}\BibitemShut {NoStop}%
\bibitem [{\citenamefont {Johnson}(2007)}]{johnson2007atomic}%
  \BibitemOpen
  \bibfield  {author} {\bibinfo {author} {\bibfnamefont {W.~R.}\ \bibnamefont {Johnson}},\ }\href@noop {} {\emph {\bibinfo {title} {Atomic Structure Theory: Lectures on Atomic Physics}}}\ (\bibinfo  {publisher} {Springer},\ \bibinfo {address} {Berlin; London},\ \bibinfo {year} {2007})\BibitemShut {NoStop}%
\bibitem [{\citenamefont {Kahl}\ and\ \citenamefont {Berengut}(2019)}]{Kahl2019}%
  \BibitemOpen
  \bibfield  {author} {\bibinfo {author} {\bibfnamefont {E.}~\bibnamefont {Kahl}}\ and\ \bibinfo {author} {\bibfnamefont {J.}~\bibnamefont {Berengut}},\ }\bibfield  {title} {\bibinfo {title} {{AMB}i{T}: A programme for high-precision relativistic atomic structure calculations},\ }\href {https://doi.org/10.1016/j.cpc.2018.12.014} {\bibfield  {journal} {\bibinfo  {journal} {Computer Physics Communications}\ }\textbf {\bibinfo {volume} {238}},\ \bibinfo {pages} {232–243} (\bibinfo {year} {2019})}\BibitemShut {NoStop}%
\bibitem [{\citenamefont {Geddes}\ \emph {et~al.}(2018)\citenamefont {Geddes}, \citenamefont {Czapski}, \citenamefont {Kahl},\ and\ \citenamefont {Berengut}}]{Geddes2018}%
  \BibitemOpen
  \bibfield  {author} {\bibinfo {author} {\bibfnamefont {A.~J.}\ \bibnamefont {Geddes}}, \bibinfo {author} {\bibfnamefont {D.~A.}\ \bibnamefont {Czapski}}, \bibinfo {author} {\bibfnamefont {E.~V.}\ \bibnamefont {Kahl}},\ and\ \bibinfo {author} {\bibfnamefont {J.~C.}\ \bibnamefont {Berengut}},\ }\bibfield  {title} {\bibinfo {title} {Saturated-configuration-interaction calculations for five-valent {T}a and {D}b},\ }\href {https://doi.org/10.1103/physreva.98.042508} {\bibfield  {journal} {\bibinfo  {journal} {Physical Review A}\ }\textbf {\bibinfo {volume} {98}},\ \bibinfo {pages} {042508} (\bibinfo {year} {2018})}\BibitemShut {NoStop}%
\bibitem [{\citenamefont {Burns}\ and\ \citenamefont {Adams}(1956)}]{Burns1956}%
  \BibitemOpen
  \bibfield  {author} {\bibinfo {author} {\bibfnamefont {K.}~\bibnamefont {Burns}}\ and\ \bibinfo {author} {\bibfnamefont {K.~B.}\ \bibnamefont {Adams}},\ }\bibfield  {title} {\bibinfo {title} {Energy levels and wavelengths of natural cadmium and of cadmium-114},\ }\href {https://doi.org/10.1364/josa.46.000094} {\bibfield  {journal} {\bibinfo  {journal} {Journal of the Optical Society of America}\ }\textbf {\bibinfo {volume} {46}},\ \bibinfo {pages} {000094} (\bibinfo {year} {1956})}\BibitemShut {NoStop}%
\bibitem [{\citenamefont {Brown}\ \emph {et~al.}(1975)\citenamefont {Brown}, \citenamefont {Tilford},\ and\ \citenamefont {Ginter}}]{Brown1975}%
  \BibitemOpen
  \bibfield  {author} {\bibinfo {author} {\bibfnamefont {C.~M.}\ \bibnamefont {Brown}}, \bibinfo {author} {\bibfnamefont {S.~G.}\ \bibnamefont {Tilford}},\ and\ \bibinfo {author} {\bibfnamefont {M.~L.}\ \bibnamefont {Ginter}},\ }\bibfield  {title} {\bibinfo {title} {Absorption spectra of {Z}n {I} and {C}d {I} in the 1300–1750 Å region},\ }\href {https://doi.org/10.1364/josa.65.001404} {\bibfield  {journal} {\bibinfo  {journal} {Journal of the Optical Society of America}\ }\textbf {\bibinfo {volume} {65}},\ \bibinfo {pages} {001404} (\bibinfo {year} {1975})}\BibitemShut {NoStop}%
\bibitem [{\citenamefont {Vidolova-Angelova}\ \emph {et~al.}(1996)\citenamefont {Vidolova-Angelova}, \citenamefont {Baharis}, \citenamefont {Roupakas},\ and\ \citenamefont {Kompitsas}}]{VidolovaAngelova1996}%
  \BibitemOpen
  \bibfield  {author} {\bibinfo {author} {\bibfnamefont {E.}~\bibnamefont {Vidolova-Angelova}}, \bibinfo {author} {\bibfnamefont {C.}~\bibnamefont {Baharis}}, \bibinfo {author} {\bibfnamefont {G.}~\bibnamefont {Roupakas}},\ and\ \bibinfo {author} {\bibfnamefont {M.}~\bibnamefont {Kompitsas}},\ }\bibfield  {title} {\bibinfo {title} {Observations and theoretical analysis of highly excited singlet and triplet states of cadmium},\ }\href {https://doi.org/10.1088/0953-4075/29/12/010} {\bibfield  {journal} {\bibinfo  {journal} {Journal of Physics B: Atomic, Molecular and Optical Physics}\ }\textbf {\bibinfo {volume} {29}},\ \bibinfo {pages} {2453} (\bibinfo {year} {1996})}\BibitemShut {NoStop}%
\bibitem [{\citenamefont {Moore}(1971)}]{Moore1971}%
  \BibitemOpen
  \bibfield  {author} {\bibinfo {author} {\bibfnamefont {C.~E.}\ \bibnamefont {Moore}},\ }\href {https://doi.org/10.6028/NBS.NSRDS.35v3} {\emph {\bibinfo {title} {Atomic Energy Levels as Derived from the Analyses of Optical Spectra: Molybdenum through Lanthanum and Hafnium through Actinium}}},\ \bibinfo {series} {National Standard Reference Data Series}, Vol.\ \bibinfo {volume} {35, Vol. III}\ (\bibinfo  {publisher} {National Bureau of Standards, U.S.},\ \bibinfo {address} {Washington, D.C.},\ \bibinfo {year} {1971})\ p.\ \bibinfo {pages} {245},\ \bibinfo {note} {reprint of NBS Circular 467, Vol. III (1958)}\BibitemShut {NoStop}%
\bibitem [{\citenamefont {Kramida}\ \emph {et~al.}(2024)\citenamefont {Kramida}, \citenamefont {{Yu.~Ralchenko}}, \citenamefont {Reader},\ and\ \citenamefont {{NIST ASD Team}}}]{NIST_ASD}%
  \BibitemOpen
  \bibfield  {author} {\bibinfo {author} {\bibfnamefont {A.}~\bibnamefont {Kramida}}, \bibinfo {author} {\bibnamefont {{Yu.~Ralchenko}}}, \bibinfo {author} {\bibfnamefont {J.}~\bibnamefont {Reader}},\ and\ \bibinfo {author} {\bibnamefont {{NIST ASD Team}}},\ }\href@noop {} {}\bibinfo {howpublished} {{NIST Atomic Spectra Database (ver. 5.12), [Online]. Available: {\tt{https://physics.nist.gov/asd}} [2025, August 17]. National Institute of Standards and Technology, Gaithersburg, MD.}} (\bibinfo {year} {2024})\BibitemShut {NoStop}%
\bibitem [{\citenamefont {Sobelman}(1979)}]{Sobelman1979}%
  \BibitemOpen
  \bibfield  {author} {\bibinfo {author} {\bibfnamefont {I.~I.}\ \bibnamefont {Sobelman}},\ }\href {https://doi.org/10.1007/978-3-662-05905-0} {\emph {\bibinfo {title} {Atomic Spectra and Radiative Transitions}}}\ (\bibinfo  {publisher} {Springer Berlin Heidelberg},\ \bibinfo {year} {1979})\BibitemShut {NoStop}%
\bibitem [{\citenamefont {Harris}\ and\ \citenamefont {Bertolucci}(1978)}]{harris1978symmetry}%
  \BibitemOpen
  \bibfield  {author} {\bibinfo {author} {\bibfnamefont {D.~C.}\ \bibnamefont {Harris}}\ and\ \bibinfo {author} {\bibfnamefont {M.~D.}\ \bibnamefont {Bertolucci}},\ }\href@noop {} {\emph {\bibinfo {title} {Symmetry and Spectroscopy: An Introduction to Vibrational and Electronic Spectroscopy}}}\ (\bibinfo  {publisher} {Oxford University Press},\ \bibinfo {address} {New York},\ \bibinfo {year} {1978})\BibitemShut {NoStop}%
\bibitem [{\citenamefont {Foot}(2005)}]{foot2005atomic}%
  \BibitemOpen
  \bibfield  {author} {\bibinfo {author} {\bibfnamefont {C.~J.}\ \bibnamefont {Foot}},\ }\href@noop {} {\emph {\bibinfo {title} {Atomic Physics}}},\ Oxford Master Series in Physics\ (\bibinfo  {publisher} {Oxford University Press},\ \bibinfo {address} {Oxford},\ \bibinfo {year} {2005})\BibitemShut {NoStop}%
\bibitem [{\citenamefont {of~Mathematical~Functions}(2025)}]{NIST_DLMF}%
  \BibitemOpen
  \bibfield  {author} {\bibinfo {author} {\bibfnamefont {N.~D.~L.}\ \bibnamefont {of~Mathematical~Functions}},\ }\href@noop {} {\bibinfo {title} {Spherical bessel functions, section 10.52}},\ \bibinfo {howpublished} {\url{https://dlmf.nist.gov/10.52}} (\bibinfo {year} {2025}),\ \bibinfo {note} {release 1.2.1 of 2025-03-15}\BibitemShut {NoStop}%
\bibitem [{\citenamefont {Abramowitz}\ and\ \citenamefont {Stegun}(1964)}]{abramowitz_stegun_1964}%
  \BibitemOpen
  \bibfield  {author} {\bibinfo {author} {\bibfnamefont {M.}~\bibnamefont {Abramowitz}}\ and\ \bibinfo {author} {\bibfnamefont {A.}~\bibnamefont {Stegun}, \bibfnamefont {Irene}},\ }\href@noop {} {\emph {\bibinfo {title} {Handbook of Mathematical Functions with Formulas, Graphs, and Mathematical Tables}}},\ Applied Mathematics Series, No. 55\ (\bibinfo  {publisher} {National Bureau of Standards},\ \bibinfo {year} {1964})\BibitemShut {NoStop}%
\bibitem [{\citenamefont {Atkins}\ and\ \citenamefont {Friedman}(2010)}]{AtkinsFriedman2010}%
  \BibitemOpen
  \bibfield  {author} {\bibinfo {author} {\bibfnamefont {P.~W.}\ \bibnamefont {Atkins}}\ and\ \bibinfo {author} {\bibfnamefont {R.~S.}\ \bibnamefont {Friedman}},\ }\href@noop {} {{\selectlanguage {English}\emph {\bibinfo {title} {Molecular Quantum Mechanics}}}},\ \bibinfo {edition} {5th}\ ed.\ (\bibinfo  {publisher} {Oxford University Press},\ \bibinfo {address} {Oxford, UK},\ \bibinfo {year} {2010})\ p.\ \bibinfo {pages} {560}\BibitemShut {NoStop}%
\bibitem [{\citenamefont {Drake}(1996)}]{Drake1996}%
  \BibitemOpen
  \bibinfo {editor} {\bibfnamefont {G.~W.~F.}\ \bibnamefont {Drake}},\ ed.,\ \href@noop {} {\emph {\bibinfo {title} {Atomic, Molecular, $\&$ Optical Physics Handbook}}}\ (\bibinfo  {publisher} {American Institute of Physics},\ \bibinfo {address} {Woodbury, NY},\ \bibinfo {year} {1996})\BibitemShut {NoStop}%
\bibitem [{\citenamefont {Ferrari}\ \emph {et~al.}(2006)\citenamefont {Ferrari}, \citenamefont {Poli}, \citenamefont {Sorrentino},\ and\ \citenamefont {Tino}}]{Ferrari2006}%
  \BibitemOpen
  \bibfield  {author} {\bibinfo {author} {\bibfnamefont {G.}~\bibnamefont {Ferrari}}, \bibinfo {author} {\bibfnamefont {N.}~\bibnamefont {Poli}}, \bibinfo {author} {\bibfnamefont {F.}~\bibnamefont {Sorrentino}},\ and\ \bibinfo {author} {\bibfnamefont {G.~M.}\ \bibnamefont {Tino}},\ }\bibfield  {title} {\bibinfo {title} {Long-lived bloch oscillations with bosonic {S}r atoms and application to gravity measurement at the micrometer scale},\ }\href {https://doi.org/10.1103/physrevlett.97.060402} {\bibfield  {journal} {\bibinfo  {journal} {Physical Review Letters}\ }\textbf {\bibinfo {volume} {97}},\ \bibinfo {pages} {060402} (\bibinfo {year} {2006})}\BibitemShut {NoStop}%
\bibitem [{\citenamefont {Santra}\ \emph {et~al.}(2004)\citenamefont {Santra}, \citenamefont {Christ},\ and\ \citenamefont {Greene}}]{Santra2004}%
  \BibitemOpen
  \bibfield  {author} {\bibinfo {author} {\bibfnamefont {R.}~\bibnamefont {Santra}}, \bibinfo {author} {\bibfnamefont {K.~V.}\ \bibnamefont {Christ}},\ and\ \bibinfo {author} {\bibfnamefont {C.~H.}\ \bibnamefont {Greene}},\ }\bibfield  {title} {\bibinfo {title} {Properties of metastable alkaline-earth-metal atoms calculated using an accurate effective core potential},\ }\href {https://doi.org/10.1103/physreva.69.042510} {\bibfield  {journal} {\bibinfo  {journal} {Physical Review A}\ }\textbf {\bibinfo {volume} {69}},\ \bibinfo {pages} {042510} (\bibinfo {year} {2004})}\BibitemShut {NoStop}%
\bibitem [{\citenamefont {Trautmann}\ \emph {et~al.}(2023)\citenamefont {Trautmann}, \citenamefont {Yankelev}, \citenamefont {Kl\"{u}sener}, \citenamefont {Park}, \citenamefont {Bloch},\ and\ \citenamefont {Blatt}}]{Trautmann2023}%
  \BibitemOpen
  \bibfield  {author} {\bibinfo {author} {\bibfnamefont {J.}~\bibnamefont {Trautmann}}, \bibinfo {author} {\bibfnamefont {D.}~\bibnamefont {Yankelev}}, \bibinfo {author} {\bibfnamefont {V.}~\bibnamefont {Kl\"{u}sener}}, \bibinfo {author} {\bibfnamefont {A.~J.}\ \bibnamefont {Park}}, \bibinfo {author} {\bibfnamefont {I.}~\bibnamefont {Bloch}},\ and\ \bibinfo {author} {\bibfnamefont {S.}~\bibnamefont {Blatt}},\ }\bibfield  {title} {\bibinfo {title} {$^1${S}$_0$ - $^3${P}$_2$ magnetic quadrupole transition in neutral strontium},\ }\href {https://doi.org/10.1103/physrevresearch.5.013219} {\bibfield  {journal} {\bibinfo  {journal} {Physical Review Research}\ }\textbf {\bibinfo {volume} {5}},\ \bibinfo {pages} {013219} (\bibinfo {year} {2023})}\BibitemShut {NoStop}%
\bibitem [{\citenamefont {Craig}\ and\ \citenamefont {Thirunamachandran}(1998)}]{Craig1998MQED}%
  \BibitemOpen
  \bibfield  {author} {\bibinfo {author} {\bibfnamefont {D.~P.}\ \bibnamefont {Craig}}\ and\ \bibinfo {author} {\bibfnamefont {T.}~\bibnamefont {Thirunamachandran}},\ }\href@noop {} {\emph {\bibinfo {title} {Molecular Quantum Electrodynamics}}}\ (\bibinfo  {publisher} {Dover},\ \bibinfo {address} {Mineola, NY},\ \bibinfo {year} {1998})\BibitemShut {NoStop}%
\bibitem [{\citenamefont {Porsev}\ and\ \citenamefont {Derevianko}(2004)}]{Porsev2004}%
  \BibitemOpen
  \bibfield  {author} {\bibinfo {author} {\bibfnamefont {S.~G.}\ \bibnamefont {Porsev}}\ and\ \bibinfo {author} {\bibfnamefont {A.}~\bibnamefont {Derevianko}},\ }\bibfield  {title} {\bibinfo {title} {Hyperfine quenching of the metastable $^3${P}$_{0,2}$ states in divalent atoms},\ }\href {https://doi.org/10.1103/physreva.69.042506} {\bibfield  {journal} {\bibinfo  {journal} {Physical Review A}\ }\textbf {\bibinfo {volume} {69}},\ \bibinfo {pages} {042506} (\bibinfo {year} {2004})}\BibitemShut {NoStop}%
\bibitem [{\citenamefont {Weiner}\ \emph {et~al.}(1999)\citenamefont {Weiner}, \citenamefont {Bagnato}, \citenamefont {Zilio},\ and\ \citenamefont {Julienne}}]{Weiner1999}%
  \BibitemOpen
  \bibfield  {author} {\bibinfo {author} {\bibfnamefont {J.}~\bibnamefont {Weiner}}, \bibinfo {author} {\bibfnamefont {V.~S.}\ \bibnamefont {Bagnato}}, \bibinfo {author} {\bibfnamefont {S.}~\bibnamefont {Zilio}},\ and\ \bibinfo {author} {\bibfnamefont {P.~S.}\ \bibnamefont {Julienne}},\ }\bibfield  {title} {\bibinfo {title} {Experiments and theory in cold and ultracold collisions},\ }\href {https://doi.org/10.1103/revmodphys.71.1} {\bibfield  {journal} {\bibinfo  {journal} {Reviews of Modern Physics}\ }\textbf {\bibinfo {volume} {71}},\ \bibinfo {pages} {1–85} (\bibinfo {year} {1999})}\BibitemShut {NoStop}%
\bibitem [{\citenamefont {Dalfovo}\ \emph {et~al.}(1999)\citenamefont {Dalfovo}, \citenamefont {Giorgini}, \citenamefont {Pitaevskii},\ and\ \citenamefont {Stringari}}]{Dalfovo1999}%
  \BibitemOpen
  \bibfield  {author} {\bibinfo {author} {\bibfnamefont {F.}~\bibnamefont {Dalfovo}}, \bibinfo {author} {\bibfnamefont {S.}~\bibnamefont {Giorgini}}, \bibinfo {author} {\bibfnamefont {L.~P.}\ \bibnamefont {Pitaevskii}},\ and\ \bibinfo {author} {\bibfnamefont {S.}~\bibnamefont {Stringari}},\ }\bibfield  {title} {\bibinfo {title} {Theory of {B}ose-{E}instein condensation in trapped gases},\ }\href {https://doi.org/10.1103/revmodphys.71.463} {\bibfield  {journal} {\bibinfo  {journal} {Reviews of Modern Physics}\ }\textbf {\bibinfo {volume} {71}},\ \bibinfo {pages} {463–512} (\bibinfo {year} {1999})}\BibitemShut {NoStop}%
\bibitem [{\citenamefont {Gribakin}\ and\ \citenamefont {Flambaum}(1993)}]{Gribakin1993}%
  \BibitemOpen
  \bibfield  {author} {\bibinfo {author} {\bibfnamefont {G.~F.}\ \bibnamefont {Gribakin}}\ and\ \bibinfo {author} {\bibfnamefont {V.~V.}\ \bibnamefont {Flambaum}},\ }\bibfield  {title} {\bibinfo {title} {Calculation of the scattering length in atomic collisions using the semiclassical approximation},\ }\href {https://doi.org/10.1103/physreva.48.546} {\bibfield  {journal} {\bibinfo  {journal} {Physical Review A}\ }\textbf {\bibinfo {volume} {48}},\ \bibinfo {pages} {546–553} (\bibinfo {year} {1993})}\BibitemShut {NoStop}%
\bibitem [{\citenamefont {Margenau}(1939)}]{Margenau1939}%
  \BibitemOpen
  \bibfield  {author} {\bibinfo {author} {\bibfnamefont {H.}~\bibnamefont {Margenau}},\ }\bibfield  {title} {\bibinfo {title} {Van der {W}aals forces},\ }\href {https://doi.org/10.1103/revmodphys.11.1} {\bibfield  {journal} {\bibinfo  {journal} {Reviews of Modern Physics}\ }\textbf {\bibinfo {volume} {11}},\ \bibinfo {pages} {1–35} (\bibinfo {year} {1939})}\BibitemShut {NoStop}%
\bibitem [{\citenamefont {O’Carroll}\ and\ \citenamefont {Sucher}(1968)}]{OCarroll1968}%
  \BibitemOpen
  \bibfield  {author} {\bibinfo {author} {\bibfnamefont {M.}~\bibnamefont {O’Carroll}}\ and\ \bibinfo {author} {\bibfnamefont {J.}~\bibnamefont {Sucher}},\ }\bibfield  {title} {\bibinfo {title} {Exact computation of the van der {W}aals constant for two hydrogen atoms},\ }\href {https://doi.org/10.1103/physrevlett.21.1143} {\bibfield  {journal} {\bibinfo  {journal} {Physical Review Letters}\ }\textbf {\bibinfo {volume} {21}},\ \bibinfo {pages} {1143–1146} (\bibinfo {year} {1968})}\BibitemShut {NoStop}%
\bibitem [{\citenamefont {Chang}(1967)}]{CHANG1967}%
  \BibitemOpen
  \bibfield  {author} {\bibinfo {author} {\bibfnamefont {T.~Y.}\ \bibnamefont {Chang}},\ }\bibfield  {title} {\bibinfo {title} {Moderately long-range interatomic forces},\ }\href {https://doi.org/10.1103/revmodphys.39.911} {\bibfield  {journal} {\bibinfo  {journal} {Reviews of Modern Physics}\ }\textbf {\bibinfo {volume} {39}},\ \bibinfo {pages} {911–942} (\bibinfo {year} {1967})}\BibitemShut {NoStop}%
\bibitem [{\citenamefont {Goebel}\ and\ \citenamefont {Hohm}(1995)}]{Goebel1995}%
  \BibitemOpen
  \bibfield  {author} {\bibinfo {author} {\bibfnamefont {D.}~\bibnamefont {Goebel}}\ and\ \bibinfo {author} {\bibfnamefont {U.}~\bibnamefont {Hohm}},\ }\bibfield  {title} {\bibinfo {title} {Dispersion of the refractive index of cadmium vapor and the dipole polarizability of the atomic cadmium $^1${S}$_0$ state},\ }\href {https://doi.org/10.1103/physreva.52.3691} {\bibfield  {journal} {\bibinfo  {journal} {Physical Review A}\ }\textbf {\bibinfo {volume} {52}},\ \bibinfo {pages} {3691–3694} (\bibinfo {year} {1995})}\BibitemShut {NoStop}%
\bibitem [{\citenamefont {Koperski}(2002)}]{Koperski2002}%
  \BibitemOpen
  \bibfield  {author} {\bibinfo {author} {\bibfnamefont {J.}~\bibnamefont {Koperski}},\ }\bibfield  {title} {\bibinfo {title} {Study of diatomic van der {W}aals complexes in supersonic beams},\ }\href {https://doi.org/10.1016/s0370-1573(02)00200-4} {\bibfield  {journal} {\bibinfo  {journal} {Physics Reports}\ }\textbf {\bibinfo {volume} {369}},\ \bibinfo {pages} {177–326} (\bibinfo {year} {2002})}\BibitemShut {NoStop}%
\bibitem [{\citenamefont {Pahl}\ \emph {et~al.}(2011)\citenamefont {Pahl}, \citenamefont {Figgen}, \citenamefont {Borschevsky}, \citenamefont {Peterson},\ and\ \citenamefont {Schwerdtfeger}}]{Pahl2011}%
  \BibitemOpen
  \bibfield  {author} {\bibinfo {author} {\bibfnamefont {E.}~\bibnamefont {Pahl}}, \bibinfo {author} {\bibfnamefont {D.}~\bibnamefont {Figgen}}, \bibinfo {author} {\bibfnamefont {A.}~\bibnamefont {Borschevsky}}, \bibinfo {author} {\bibfnamefont {K.~A.}\ \bibnamefont {Peterson}},\ and\ \bibinfo {author} {\bibfnamefont {P.}~\bibnamefont {Schwerdtfeger}},\ }\bibfield  {title} {\bibinfo {title} {Accurate potential energy curves for the group 12 dimers {Z}n$_2$, {C}d$_2$, and {H}g$_2$},\ }\href {https://doi.org/10.1007/s00214-011-0912-1} {\bibfield  {journal} {\bibinfo  {journal} {Theoretical Chemistry Accounts}\ }\textbf {\bibinfo {volume} {129}},\ \bibinfo {pages} {651–656} (\bibinfo {year} {2011})}\BibitemShut {NoStop}%
\bibitem [{\citenamefont {Qiao}\ \emph {et~al.}(2012)\citenamefont {Qiao}, \citenamefont {Li},\ and\ \citenamefont {Tang}}]{Qiao2012}%
  \BibitemOpen
  \bibfield  {author} {\bibinfo {author} {\bibfnamefont {L.~W.}\ \bibnamefont {Qiao}}, \bibinfo {author} {\bibfnamefont {P.}~\bibnamefont {Li}},\ and\ \bibinfo {author} {\bibfnamefont {K.~T.}\ \bibnamefont {Tang}},\ }\bibfield  {title} {\bibinfo {title} {Dynamic polarizabilities of {Z}n and {C}d and dispersion coefficients involving group 12 atoms},\ }\href {https://doi.org/10.1063/1.4746155} {\bibfield  {journal} {\bibinfo  {journal} {The Journal of Chemical Physics}\ }\textbf {\bibinfo {volume} {137}},\ \bibinfo {pages} {084309} (\bibinfo {year} {2012})}\BibitemShut {NoStop}%
\bibitem [{dat()}]{data}%
  \BibitemOpen
  \href {https://github.com/DrPaulRobert/Cadmium-AMBiT} {\bibinfo {title} {https://github.com/drpaulrobert/cadmium-ambit}}\BibitemShut {NoStop}%
\end{thebibliography}%
\end{document}